\documentclass[%
 reprint,
 amsmath,amssymb,
 aps,
]{revtex4-2}

\usepackage{graphicx}
\usepackage{dcolumn}
\usepackage{bm}
\usepackage{hyperref}

\begin{document}

\preprint{APS/123-QED}

\title{Dynamics of wire frame glasses in two dimensions}

\author{David A. King}
 \email{daviking@sas.upenn.edu}
\affiliation{Department of Physics and Astronomy, University of Pennsylvania, 209 South 33rd St., Philadelphia, PA, 19104.
}%
\date{\today}

\begin{abstract}
  The dynamics of wire frame particles in concentrated suspension are studied by means of a 2D model and compared to those of rod-like particles. The wire frames have bent or branched structures constructed from infinitely thin rigid rods. In the model, a particle is surrounded by diffusing points that it cannot cross. We derive a formal expression for the mean squared displacement (MSD) and, by using a self-consistent approximation, we find markedly different dynamics for wire frames and rods. For wire frames, there exists a critical concentration of points above which they become frozen with the long time MSD reaching a plateau. Rods, on the other hand, always diffuse by reptation. We also study the rheology through the elastic stress, and more striking differences are found;  the initial magnitude of the stress for wire frames is much larger than for rods, scaling like the \textit{square} of the point concentration, and above the critical concentration, the stress for wire frames appears to persist indefinitely while for rods it always decays.
\end{abstract}
\maketitle
\onecolumngrid
	\section{Introduction}
	Increasing the concentration of Brownian particles in suspension can have a profound effect on their dynamics, even for simple hard spheres. In particular, the mean squared displacement of the particles changes from Einstein's celebrated result for spheres in a viscous fluid\cite{Einstein1905UberTeilchen, Dhont1996AnColloids,Leegwater1992Long-Fluids, Leegwater1992DynamicalSuspensions, Szamel1992Long-timeSuspensions}. Einstein also showed that even a dilute suspension of spheres affects the rheology; the response remains Newtonian but the viscosity increases \cite{Einstein1906EineMolekul-dimensionen, Einstein1911BerichtigungMolekul-dimensionen}. However, at the next order in concentration, Batchelor found that the response becomes non-Newtonian \cite{Batchelor1972TheC2}. If the concentration is increased still further, then the system will ultimately jam and no longer flow \cite{Trappe2001}.
		
	Other more complicated kinds of suspended particles can be considered. Rigid particles of general shapes are a natural extension \cite{Happel1973LowMedia.,Kim2005MicrohydrodynamicsApplications,Makino2004BrownianFluid,Makino2004ViscoelasticityShape} with many realisations; the tobacco mosaic virus is an example of an (approximately) rigid rod \cite{Dubin1967,Nemoto1975}, Laponite clays are flat disks \cite{Nicolai2000} and DNA origami techniques can be used to design any desired shape \cite{Seeman1982NucleicLatticesb,Rothemund2006FoldingPatterns,Han2011DNASpace, Castro2011AOrigami}. In the same way as for Brownian spheres, the various motions of these particles which are well understood in dilution, are strongly modified in concentrated suspension. The rheology of these suspensions also follows the same pattern, and a wide range of non-Newtonian flow behaviours are seen at higher concentrations. It is this diversity of dynamics and mechanical responses that make these systems so technologically important and fundamentally interesting. One key question is how to predict these behaviours based on the shape of the particles and their dominant interactions. 
 
	It is particularly interesting to consider suspensions of very long thin particles. In general, even at very low volume fractions, the dynamics of these systems are very slow and their rheology significantly non-Newtonian. This is true despite the fact that, when the particles are sufficiently long compared to their width, the excluded volume interactions are un-important. In the absence of any other interactions, this makes the equilibrium behaviour identical to the ideal gas. The culprit for this behaviour is the constraint that no two particles may cross through each other. These are ``kinetic'' or ``topological'' constraints \cite{Edwards1967StatisticalI, Edwards1967StatisticalII, Doi1986TheDynamics}. In this paper, we focus on how the geometry of these highly extended, rigid particles affects their dynamics and rheology. These questions have a long history so, before outlining our work in detail, it is worth placing our work in context.
	
	Perhaps the simplest example of such a system is a concentrated suspension of thin rigid rods. If the rods are of length $L$, diameter $b$ and number density $\rho$, then by concentrated we mean $\rho L^3 \gg 1$, but because the rods are thin we have $\rho L^2 b \ll 1$. This second limit ensures that the excluded volume interactions are indeed negligible and no orientational order is present in the system \cite{Doi1986TheDynamics,Onsager1949TheParticles}. The dynamics of this system were successfully understood using two cornerstones of modern polymer theory; the ``tube'' and ``reptation'' \cite{Doi1975RotationalSolution,Doi1978Dynamics1,Doi1978Dynamics2,DeGennes1971ReptationObstacles}. One of the key results from this approach is that the rotational diffusion of the rods, as well as that transverse to their length, is strongly suppressed by a factor $(\rho L^3)^{-2}$ \cite{Doi1975RotationalSolution, Doi1978Dynamics1}. This ultimately leads to pronounced non-Newtonian rheology, with the ratio between the steady state elastic and viscous shear stresses scaling $\sim (\rho L^3)^2$ \cite{Doi1978Dynamics2, Doi1986TheDynamics}.
	
	To investigate how these properties depend on the shape of the particle we consider generalising the rod to a ``wire frame''. These are particles constructed from rods to make bent or branched structures. Examples of these have been realised using DNA origami to build branched Y-shaped wire frames or more general ``nano-stars'', where each leg is made from double stranded DNA \cite{Xing2018MicrorheologyHydrogels, Biffi2013PhaseNanostars}. The long persistence length $\sim 390$\r{A} and high aspect ratio $\sim 20$ of the double stranded DNA legs mean these particles are well approximated by perfectly thin and rigid theoretical idealisations \cite{Gross2011QuantifyingTension}. 
	
	Another example of wire frame particles which have been studied are ``Onsager crosses'', constructed from three mutually perpendicularly bisecting rods of equal length. Their dynamics are very rich \cite{VanKetel2005StructuralGas}; simulations of these particles with fixed orientations show that the system has the features of a strong glass former, with the translational diffusion of the particles becoming extremely slow. In particular, the diffusion constant is found to decrease exponentially $D \sim e^{-a \rho L^3}$, with $a$ a positive constant. Their rheology has also been considered, with simulations showing that the viscosity of the suspension is very large and sensitive to concentration \cite{Heine2010EffectSuspensions,Petersen2010ShearNanoparticles}. Recently, this was addressed theoretically, using a simple geometric approach to determine the scaling of the initial elastic shear stress with density \cite{King2021ElasticModelc,King2021ElasticSystemsc}. That work found results consistent with simulations, where the ratio between the elastic response of wire frames and rods scales $\sim \rho^2$. It was also observed that this behaviour was highly dependent on the particle shape, with even a particle bent through a very slight angle, of say $15^{\circ}$, behaves radically differently from a straight rod. These results have been supported by recent experiments comparing helical filaments to straight rods \cite{Yardimci2023}.
	
	The reason for this starkly different behaviour is the lack of reptation dynamics in these systems. According to the reptation picture, the motion of a rod in concentrated suspension is facilitated by its freedom of movement along its own length. On the other hand, consider an L-shaped wire frame; there is no direction in which the particle can move before quickly becoming entangled with the surroundings. This makes it difficult to handle the dynamics without resorting to heuristic arguments, because the non-crossing constraints need to be handled exactly. Cichocki showed how this may be done by incorporating the constraints, formulated as boundary conditions, as extra forces in the Smoluchowski equation \cite{Cichocki1987TheCores}. This has then been used as the starting point in the study of a number of related systems, such as the rotation of rods fixed to a lattice \cite{Schilling2003MicroscopicCorrelations,Schilling2003MicroscopicCorrelations} and, importantly for us, the reptation of rods and the diffusion of branched cross wire frames \cite{Sussman2011MicroscopicMacromolecules}. The study of reptation in this framework by Szamel and Schweizer focused on the transverse diffusion of rods with fixed orientation \cite{Szamel1993ReptationPolymers,Szamel1994ReptationPolymers}. By means of a dynamical mean field theory, they reproduced the same suppression of the diffusion coefficient predicted by reptation theory. This was used by Sussman and Schweizer for the translational diffusion of crosses \cite{Sussman2011MicroscopicMacromolecules}, who found a complete cessation of motion ($D \to 0$) above a certain density. When they included dynamical density fluctuations, however, the exponential reduction of the diffusion coefficient found in simulations \cite{VanKetel2005StructuralGas} was recovered.  
	
	Another standard approach to studying the dynamics in glassy systems is ``mode coupling theory'' \cite{Reichman2005, Fuchs2009AFlow}. This approach has been quite successful, however it has been shown that the standard mode coupling theory cannot reproduce reptation results for rods because it does not appropriately encode the kinetic constraints \cite{Miyazaki2002}. This failure is due to the ``Gaussian approximation'', where fourth moments of density fluctuations are approximately expressed in terms of second moments. When the theory is extended beyond this approximation to include the effects of entanglements, results consistent with the dynamic mean field theory of Szamel and Schweizer \cite{Szamel1993ReptationPolymers, Szamel1994ReptationPolymers} are obtained \cite{Miyazaki2002}.
	 
	There are also alternative treatments based on the approach first put forward by Edwards and Evans to study the longitudinal diffusion of rods in highly concentrated suspension \cite{Edwards1982DynamicsMolecules}. Here, by first understanding the reflection of a test rod from one background rod, the single rod dynamics are built up using a mean field theory argument. This approach has been extended several times, starting with Edwards and Vilgis who included many particle, co-operative effects \cite{Edwards1986TheTransition}. Teraoka and Hayakawa studied the transverse and rotational diffusion of the rods, recovering scaling results consistent with reptation \cite{Teraoka1988TheoryDiffusion, Teraoka1989}. Collective effects were also studied using a more detailed method by Teraoka and Karasz \cite{Teraoka1992One-dimensionalBarriers, Teraoka1993GlassMolecules}. While this framework is appealing because it presents a very clear physical picture that allows for simple qualitative arguments to be made, the process of generalising it to different particle shapes and applying it to rheology is not straightforward. 

	In this paper we aim to compare the dynamics and rheology of concentrated suspensions of wire frame particle and rigid rods. We combine the approaches of Szamel and Schweizer and Edwards and Vilgis, and show how these may be extended to address the rheology. Due to the complex geometry of the problem, we choose to focus on a simplified two dimensional model. This model is introduced in Section \ref{sec:2DModel} and is essentially the same as that in \cite{King2021ElasticModelc}, but extended to include Brownian dynamics. It was shown that the static model produced equivalent results in two and three dimensions \cite{King2021ElasticSystemsc}, and we argue that it contains all of the essential features of a fully three dimensional system. We also only focus on one wire frame particle shape; a bent particle constructed from two rods meet at an angle $\chi$. The same general conclusions should apply to other wire frame geometries. 
	
	In Section \ref{sec:MSD}, starting from Cichocki's generalised Smoluchowski equation, we investigate the mean squared displacement of the particles in our model. We ignore the translational diffusion, since it has been shown previously to be extremely slow for wire frames, and focus only on their rotation. We write the mean squared displacement in terms of a time dependent diffusion kernel, for which we derive a general expression. This is the same as that found by Edwards and Vilgis \cite{Edwards1986TheTransition}. To probe higher concentrations more efficiently, in Section \ref{sec:MSDselfconsapp} we invoke a self consistent approximation. This is essentially that used by Szamel and Schweizer \cite{Szamel1993ReptationPolymers, Szamel1994ReptationPolymers} and mode coupling theory \cite{Reichman2005, Bouchaud1996}, although formulated differently. We find striking differences between the behaviour of rods and wire frames. While reptation allows rods to diffuse even at high concentrations, there is a critical concentration at which the motion of wire frames becomes sub-diffusive and, at higher concentrations, the mean squared displacement plateaus completely.
	
	The rheology is addressed in Section \ref{sec:Rheology}. Our primary focus is the elastic response, and we shall ignore the viscous part by considering only step strain, so that it may be taken to vanish for all practical timescales. We use the virtual work principle to derive an expression for the elastic stress including the effect of the constraints. We find results consistent previous work for the magnitude of the initial elastic stress after the step strain \cite{King2021ElasticModelc} and further show that its relaxation is much more complicated than it is for rods. At the same critical concentration marking the departure from pure diffusive behaviour, the stress relaxation changes from exponential decay to a power law. When the concentration is increased further it may persist indefinitely, although we expect that collective effects not addressed here mean that it will eventually relax. We conclude in Section \ref{sec:Conclusions} with a discussion of our results and how the work presented here may be extended.  
					
 	\section{Two Dimensional Model}
	\label{sec:2DModel}
 	\begin{figure}\includegraphics[width=8cm]{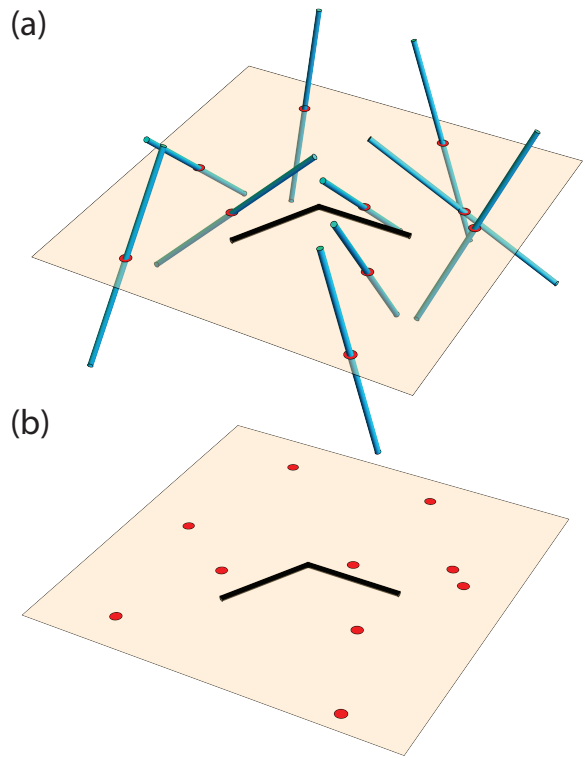}
		\caption{\label{fig:3DSlice} A sketch of how the two-dimensional model treated in this paper arises as a slice of the three-dimensional system. Panel (a) shows the 3D system; a  black bent wire frame is surrounded by the cyan legs of surrounding wire frames. The points where these legs intersect the shaded plane of the cross are shown in red. The resulting 2D system is shown in panel (b); because the wire frame cannot intersect the other legs, it cannot cross the red points where they pass through its plane.}
	\end{figure}
 
	To construct the model, we restrict our attention to only planar wire frames and imagine taking a slice through a three dimensional system in the plane of a particular particle. As shown in Fig.(\ref{fig:3DSlice}), this leads to a picture with the test particle in the plane surrounded by a collection of $N$ constraining points. These are the points where the surrounding particles intersect the plane. We shall work in the limit as $N \to \infty$ while maintaining a fixed number density $\rho = N/V$, where $V$ is the volume (or area in two dimensions) of the system. Here we only consider two types of particle; a straight rod of length $L$ and a bent wire frame formed of two such rods meeting at an angle $\chi \leq \pi$. We suppose that these rods are infinitely thin mathematical line segments and the constraints are mathematical points. Therefore, there is no excluded volume interactions but the particle and constraints still interact via the restriction that they cannot pass through each other \cite{Frenkel1983MolecularRods,Frenkel1981MolecularProperties}. This means that, while the equilibrium distribution is (almost) trivial, the dynamics are severely affected by the constraints. It is this affect on which we are focused. 
	
	The dynamics are included by allowing both the particle and the constraining points to diffuse. This supposes that the system is immersed in a viscous solvent, although, for simplicity, we shall ignore the hydrodynamic interactions between the particle and constraints. In a real three dimensional system, one can make a geometric argument relating the diffusion constant of the constraining points to the translational and rotational diffusion constants of the particles, however, again for simplicity, we ascribe a separate diffusion constant to constraints. In the absence of the particle and any other constraints, the ``free case'', a single constraining point diffuses with diffusion constant $\mathcal{D}$. For the wire frame, we shall only consider its rotational motion, since previous studies indicate that the translational motion is extremely slow \cite{VanKetel2005StructuralGas,Sussman2011MicroscopicMacromolecules}. The rotational motion also primarily controls the stress relaxation \cite{Doi1978Dynamics2,Doi1986TheDynamics}. The free rotational diffusion constant of the particle we call $D_r$. It is important to note that the two dimensional model is strictly a slice of a three dimensional system. This allows us to use familiar hydrodynamic formulae without modification. 

    At this early stage, it is important to mention some assumptions we shall employ throughout. We assume that the system is initially prepared in equilibrium with a Boltzmann distribution. This is standard and is correct when the system is ergodic, but it is only one of many possible preparation protocols for these glassy systems. 
	Another important assumption is that the system remains entirely homogenous and, when discussing the rheology, we make the same assumption for the flow rate. 
	\subsection{Governing Equations}
   \begin{figure}\includegraphics[width=8cm]{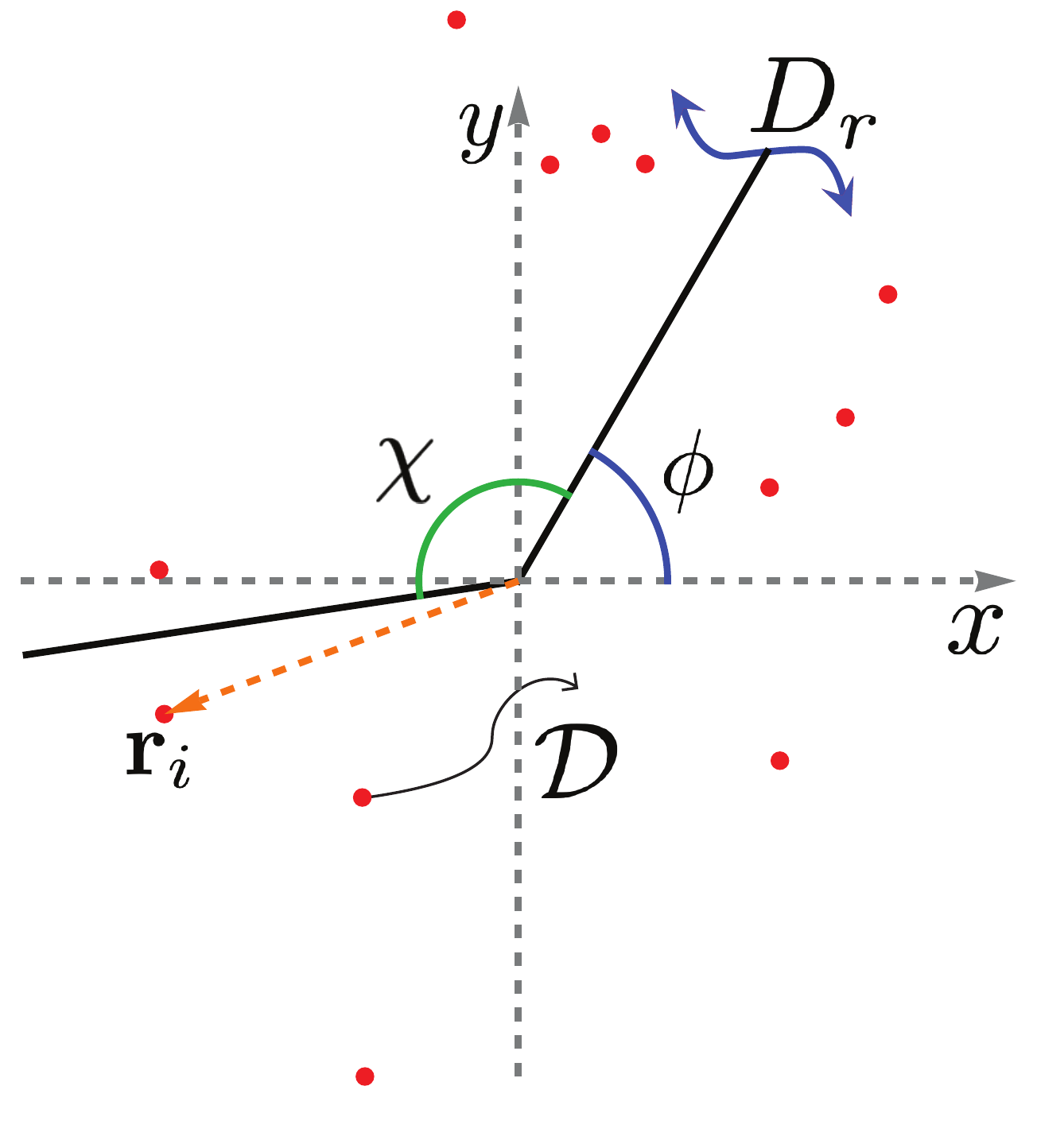}
		\caption{\label{fig:SystemSketch} A sketch of the two-dimensional system. The wire frame in the plane is surrounded by red constraining points which it cannot cross. The orientation of the wire frame is specified by the angle $\phi$ between the $x$-axis and one of its legs. Its two legs are seperated by an angle $\chi$. The positions of the constraining points are the vectors $\textbf{r}_i$. The centre of the wire frame remains fixed at the origin while it undergoes rotational diffusion with free diffusional constant $D_r$. The constraints diffuse with diffusion constant $\mathcal{D}$.}
	\end{figure}
	In the presence of the constraints we assume that the rod particle behaves according to the reptation theory \cite{Doi1975RotationalSolution,Doi1978Dynamics1,Doi1978Dynamics2,Doi1986TheDynamics}, and shall restate these well known results when relevant. For the wire frame, our aim is to understand how its dynamics are modified in the presence of the constraints, to do this we need to define its position and orientation. We suppose that the wire frame does not undergo translational diffusion, so we can take its centre to remain fixed at the origin. Its orientation is defined by the angle $\phi$ between one constituent rod and the $x$-axis. The $N$ constraints are at positions $\textbf{r}_i$, with $i \in [1,N]$ indexing the particular constraint. It is useful for us to write the constraint positions in polar co-ordinates, with radial positions $r_i$ and angles $\varphi_i$. The system is sketched in Fig.(\ref{fig:SystemSketch}) with the particle and constraint positions indicated.
    \begin{figure}\includegraphics[width=8cm]{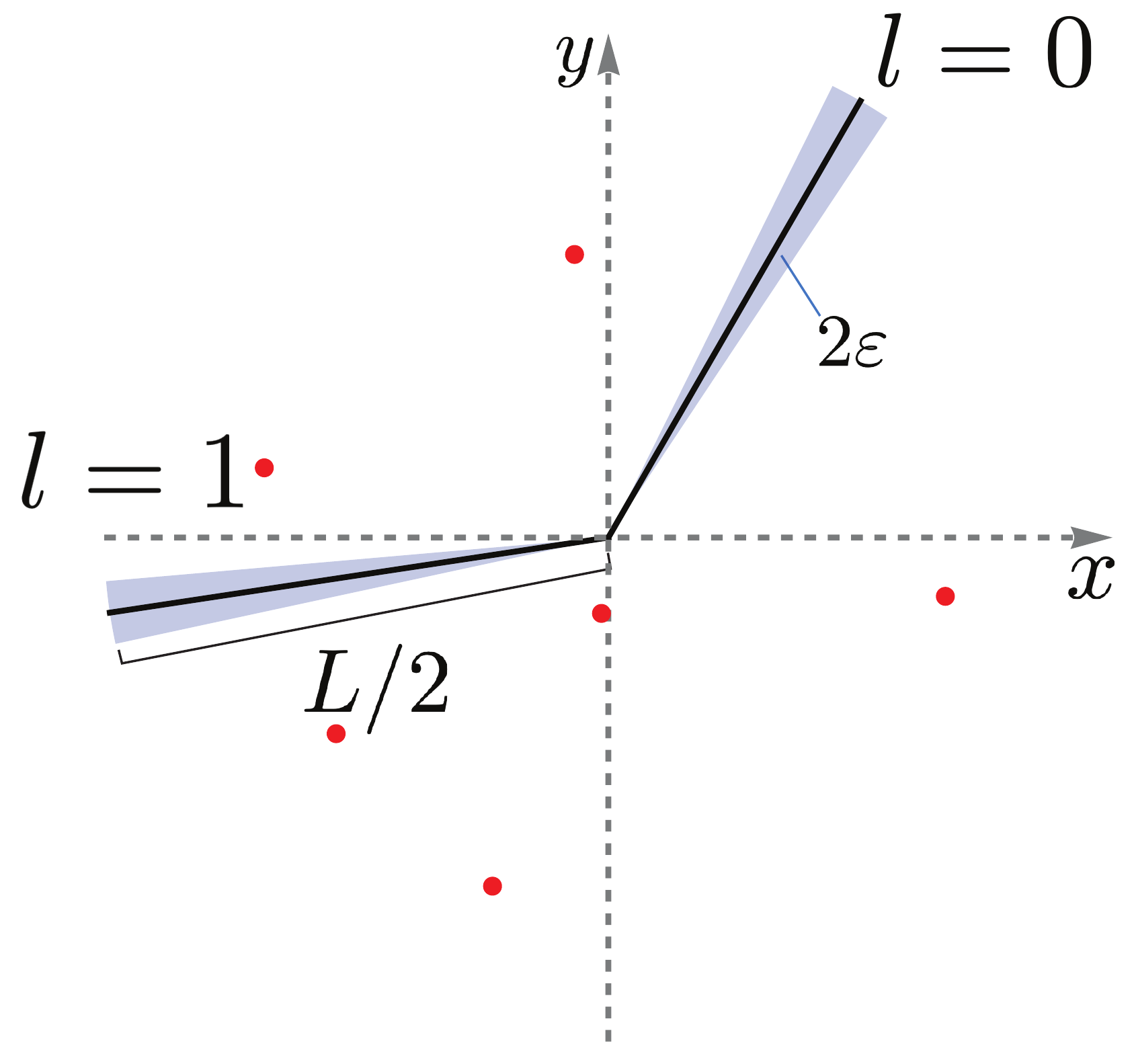}
		\caption{\label{fig:InflatedParticle} A sketch of the particle once it has been ``inflated''. Each black leg is replaced by the blue circle segments with radius $L/2$ and opening angle $2\varepsilon$. The red constraining points cannot enter these regions. Formulating these restrictions as boundary conditions in the Smoluchowski equation leads to the effective force in Eqn.(\ref{Tdefns1}).}
	\end{figure}
	
	The Smoluchowski equation which governs the evolution of the distribution function $P(\phi, \{\textbf{r}_i\} ;t)$, including the non-crossing constraints, may be derived by the method provided by Cichoki \cite{Cichocki1987TheCores}. The constraints are written as boundary conditions on the relative probability current between the particle and constraining points, requiring it to vanish whenever they collide. These can then be incorporated into the Smoluchowski equation as additional forces and torques. While the method is outlined in detail for hard spheres and rigid rods with finite diameters, it is straightforward to modify it to our case. It is easiest to first consider the case where the rods making up the particle have been inflated from mathematical lines to circle segments with radius $L/2$ and opening angle $2\varepsilon$, as shown in Fig.(\ref{fig:InflatedParticle}). The resulting effective torque on this particle due to its interaction with constraint $i$ is given by
	\begin{equation}
		\label{Tdefns1}
		\Theta_i(L/2 - r_i)\mathcal{T}^{\varepsilon}_i(\phi -\varphi_i)  = \Theta(L/2-r_i) \sum_{l =0,1} \delta(\varepsilon - |\phi - \varphi_i+l\chi|) \ \text{sgn}(\phi - \varphi_i+l\chi),
	\end{equation} 
	where $\Theta(x)$ is the Heaviside step function. The label $l$ indexes each leg of the wire frame, as shown in Fig.(\ref{fig:InflatedParticle}). This equation \textit{defines} the function $\mathcal{T}_i^{\varepsilon}$. The torque on our infinitely thin particle is then understood as a limiting case of the above
	\begin{equation}
		\label{Tdefns}
		\mathcal{T}_i \Theta_i \equiv \mathcal{T}_i(\phi -\varphi_i)\Theta_i(L/2 - r_i) = \lim_{\varepsilon \to 0} \mathcal{T}^{\varepsilon}_i(\phi -\varphi_i) \Theta_i(L/2 - r_i),
	\end{equation}
	where we define the shorthand notations $\mathcal{T}_i$ and $\Theta_i$ which we shall use throughout. Note that there is also an equal an opposite force on the constraint $i$ due to its interaction with the particle. This results in the Smoluchowski equation for the system which includes the kinetic constraints as forces,
		\begin{equation}
		\label{SEqn}
		\begin{split}
			\frac{\partial P}{\partial t} &=  D_r \frac{\partial}{\partial \phi}\left(\frac{\partial}{\partial \phi} - \sum_{i=1}^{N} \mathcal{T}_i \Theta_i \right)P + \sum_{i=1}^{N} \mathcal{D}\bm{\nabla}_i \cdot \left(\bm{\nabla}_i + \frac{1}{r_i} \mathcal{T}_i \Theta_i \hat{\bm{\phi}}\right)P.
		\end{split}
	\end{equation}
	Here, $\hat{\bm{\phi}}$ is the angular unit vector. It is crucial to note the equilibrium distribution, $P_{\text{eq}}$, implied from this equation. It must satisfy the following relations, which will prove invaluable later
	\begin{equation}
		\label{Pequseful}
		\frac{\partial P_{\text{eq}}}{\partial \phi} = \sum_{i=1}^{N} \mathcal{T}_j \Theta_j P_{\text{eq}}, \ \ \ \text{and} \ \ \  \frac{\partial P_{\text{eq}}}{\partial \varphi_i} = -\mathcal{T}_i \Theta_i P_{\text{eq}}.
	\end{equation}
 This means that we have 
 \begin{equation}
		\label{Peqformal}
		P_{\text{eq}} = \frac{1}{V^N} \lim_{\varepsilon \to 0} \left[1-\prod_{i=1}^{N} \sum_{l=0,1}\Theta\left(\varepsilon - |\phi - \varphi_i+l\chi|\right) \Theta(L/2-r_i)\right].
	\end{equation}
 In the limit, this is constant almost everywhere and we may approximate $P_{\text{eq}} \approx V^{-N}$. However, due to their singular nature, we must still keep track of its derivatives using (\ref{Pequseful}). External fields and fluid flows which may modify the equilibrium distribution can be included in this equation using standard methods, the latter will be discussed with reference to the rheology in Section \ref{sec:Rheology}.
	
The solution to the Smoluchowski equation can be used to compute expectation values; for a general function $Q(\phi , \{\textbf{r}_i\})$ we write,
	\begin{equation}
		\left\langle Q(t) \right\rangle = \int d\phi \prod_{i=1}^{N} d\textbf{r}_i \ P(\phi , \{\textbf{r}_i\};t)  \ Q(\phi, \{\textbf{r}_i\}).
	\end{equation}
	This is notationally unwieldy, so we shall often use the shorthand 
	\begin{equation}
		\label{Shorthand}
		\left\langle Q(t) \right\rangle = \int P(t) \ Q 
	\end{equation}
	\section{Mean Squared Displacement}
	\label{sec:MSD}
	\subsection{Formal Expression}
	\label{sec:MSDformalexpression}
	The fluctuation dissipation theorem relates the average angular velocity of the particle when acted on by a small external torque to the mean squared displacement in equilibrium \cite{Doi1986TheDynamics}. Supposing that the torque originates form the potential 
	\begin{equation}
		\label{FDTpot}
		U = - k_B T g \phi,
	\end{equation}
	the smallness of the force here implies the limit $g\to 0$. The Smoluchowski equation under the influence of this potential is, 
	\begin{equation}
		\label{SEqnPot}
		\begin{split}
			\frac{\partial P}{\partial t} &=  D_r \frac{\partial}{\partial \phi}\left(\frac{\partial}{\partial \phi} - \sum_{i=1}^{N} \mathcal{T}_i \Theta_i - g \right)P + \sum_{i=1}^{N} \mathcal{D}\bm{\nabla}_i \cdot \left(\bm{\nabla}_i + \frac{1}{r_i} \mathcal{T}_i \Theta_i \hat{\bm{\phi}}\right)P.
		\end{split}
	\end{equation}
	The average against the distribution, $P$, which solves this equation, we write as $\langle \cdots \rangle_{g}$. Then, the fluctuation dissipation theorem states that 
	\begin{equation}
		\label{FDT}
		\lim_{g \to 0} \frac{\partial}{\partial t}\left\langle\phi(t)\right\rangle_{g} = \frac{g}{2 k_B T} \left\langle \left(\phi(t)-\phi(0)\right)^2\right\rangle_{\text{eq}} \equiv \frac{g}{2 k_B T} \mathcal{M}(t)
	\end{equation}
	The average $\langle \cdots\rangle_{\text{eq}}$ on the right hand side is taken in equilibrium ($g = 0$) and defines the angular mean squared displacement as a function of time, $\mathcal{M}(t)$.
	
	It is much more efficient for us to work with the Laplace transform of these expressions. For instance, the Laplace transform of the mean squared displacement is defined as
	\begin{equation}
		\mathcal{M}(s) \equiv \int_{0}^{\infty} dt \ e^{-s t} \mathcal{M}(t).
	\end{equation}
	Note that we do not use a different symbol for a function and its Laplace transform; there is no ambiguity because we always indicate whether it is a function of $t$ or $s$. We can write $\mathcal{M}(t)$ in terms of a time dependent rotational diffusion kernel, $\widetilde{D}_r(t)$,
	\begin{subequations}
		\begin{equation}
			\mathcal{M}(t) = 2\int_{0}^{t} dt' \int_{0}^{t'} dt'' \ \widetilde{D}_r(t''),
		\end{equation}
		as is standard. In Laplace transform this reads
		\begin{equation}
			\label{MSDDr}
			\mathcal{M}(s) = 2\frac{\widetilde{D}_r(s)}{s^2}.
		\end{equation}
	\end{subequations}
	Due to the entanglements of the particle with the surrounding constraints, $\widetilde{D}_r(t)\neq D_r \delta(t)$ for non-zero number density $\rho \neq 0$. 
	
	The goal of this section is to derive a formal expression for the diffusion kernel as a function of time and number density. To make progress we write the fluctuation dissipation theorem in Laplace transform and re-arrange slightly
	\begin{equation}
		\label{FDT2}
		s \left\langle \phi(s)\right\rangle_{g} = \frac{g \widetilde{D}_r(s)}{k_B T s}. 
	\end{equation}
	This could equally well have been our starting point, without referencing the fluctuation dissipation theorem. This equation would then serve as a definition of $\widetilde{D}_r(s)$. Phrased in this way, the derivation is the same as has been used for Liouville systems \cite{Edwards1965ExactCoefficients}.
	
	 We can compute the right hand side of (\ref{FDT2}) by taking the moment of $\phi$ with the Laplace transform of the Smoluchowski equation (\ref{SEqnPot}). This gives (in the shorthand of equation (\ref{Shorthand})) an expression for $\widetilde{D}_r(s)$,
	\begin{equation}
		\label{FDT3}
		\frac{g}{k_B T}\widetilde{D}_r(s) = g D_r + D_r \sum_{i=1}^{N}\int \mathcal{T}_{i} \Theta_i P(s),
	\end{equation}
	where $P(s)$ is the Laplace transform of the solution to (\ref{SEqnPot}).
	
	The fluctuation dissipation theorem dictates we take the limit $g \to 0$ of this relation, hence we must solve (\ref{SEqnPot}) to first order in $g$. This is done by expanding about the equilibrium solution
	\begin{equation}
		P(s) = P_{\text{eq}} + \delta P(s), 
	\end{equation}
	and insisting $\delta P \sim \mathcal{O}(g)$. The solution for $\delta P$ is,
	\begin{equation}
		\label{SEqnSolng}
		\begin{split}
			\delta P(s) =& - g D_r\int d\phi' \prod_{i=1}^{N}d\textbf{r}'_i \ G_N\left(\phi,\{\textbf{r}_i\};\phi',\{\textbf{r}'_i\}|s\right) \frac{1}{s}\frac{\partial P_{\text{eq}}}{\partial \phi}(\phi',\{\textbf{r}'_i\})\\
			=&- g D_r \int G_N(s) \frac{1}{s} \frac{\partial P_{\text{eq}}}{\partial \phi},
		\end{split}  
	\end{equation}
	where $G_N$ is the Green function to the $N$-constraint Smoluchowski equation (\ref{SEqn}) and in the second line, the short hand of equation (\ref{Shorthand}) is employed. It also must be noted that the Laplace transform of unity is $s^{-1}$. Finally, we recall the derivative of the equilibrium distribution (\ref{Pequseful}) and we obtain
	\begin{equation}
		\widetilde{D}_r(s) = D_r  - D_r^2 \sum_{i,j=1}^{N} \int \int  \mathcal{T}_i \Theta_i G_N(s) \mathcal{T}_j \Theta_j P_{\text{eq}}. 
	\end{equation}
	One last manipulation is needed to make this expression usable; we use the adjoint of the Green function, $G_N^{\dagger}$, defined by
    \begin{equation}
     \int \int g \ G_N \ h = \int \int h \ G_N^{\dagger} \ g,
    \end{equation}
    for arbitrary functions $g$ and $h$. This allows us to remove the action of $G_N$ on the equilibrium distribution to give our final expression for the diffusion kernel
	\begin{equation}
		\label{DrFDTgen}
		\widetilde{D}_r(s) = D_r  - D_r^2 \sum_{i,j=1}^{N} \int \int P_{\text{eq}}  \mathcal{T}_i \Theta_i G_N^{\dagger}(s) \mathcal{T}_j \Theta_j .
	\end{equation}
	This is precisely the expression derived by Edwards and Vilgis, albeit modified to account for the general time dependence of the diffusion kernel. We should note that we have not included the position or orientation dependence of the diffusion kernel. In principle this can be done by a minor elaboration of notation and, if it is, then we could get more insights such into the spatial structure of the system, but this is not our main focus. 
	
	This is a general result, but it is not especially useful in its current form. It involves the full $N$-constraint dynamics through $G_{N}^{\dagger}$, which cannot be solved for exactly. There are, however, several potential ways of making progress. Perhaps the most natural would be to expand the second term in (\ref{DrFDTgen}) as a series in powers of density, $\rho$. Each term in this series would give the correction to the diffusion kernel originating from interactions of the the particle with increasingly many constraints. This is indeed the approach taken by Edwards and Vilgis \cite{Edwards1986TheTransition}. Such a series for $\widetilde{D}_r$ will work well for low concentrations, where only a few terms are needed. However, as we are interested in concentrated suspensions, it is impossible to develop this series to sufficient order. Therefore, a different argument is needed for high density systems. One possibility is to make a physical argument about which terms in the series are dominant and attempt to re-sum them, as done in \cite{Edwards1986TheTransition}. On the other hand, perhaps an easier option is to apply a self-consistent approximation along the lines of the dynamical mean field theory presented in \cite{Szamel1993ReptationPolymers, Szamel1994ReptationPolymers}. This is the approach we shall take.
	\subsection{Self-consistent One-Constraint Approximation}
	\label{sec:MSDselfconsapp}
	Our starting point is (\ref{FDT3}). The second term on the right hand side shows us how the dynamics of the particle are affected by the constraints and may be written in terms of the reduced one-constraint distribution, $P_1$, as follows
 	\begin{equation}
		\label{FDT3SC}
		\frac{g}{k_B T}\widetilde{D}_r(s) = g D_r + D_r \int d\phi d\textbf{r} \ \mathcal{T}(\phi-\varphi) \Theta(L/2-r) P_1(\phi,\textbf{r}|s). 
	\end{equation}
    The reduced one-constraint distribution is defined in the usual way by integrating over all but one of the $N$ constraints
    \begin{equation}
    \label{eqn:1ConDist}
        P_1(\phi, \textbf{r}|s)= N \int \prod_{k=2}^{N} d\textbf{r}_k P(\phi,\textbf{r},\{\textbf{r}_k\}|s).
    \end{equation}
    By integrating the Smoluchowski equation (\ref{SEqnPot}) in the same way, we find that $P_1$ satisfies
  	\begin{equation}
		\label{SEqnPot1}
		\begin{split}
			sP_1(\phi,\textbf{r}|s) =&  D_r \frac{\partial}{\partial \phi}\left(\frac{\partial}{\partial \phi} - \mathcal{T}(\phi-\varphi) \Theta(L/2-r) - g \right)P_1(\phi,\textbf{r}|s)\\
   &+ \mathcal{D}\bm{\nabla} \cdot \left(\bm{\nabla} + \frac{1}{r} \mathcal{T}(\phi-\varphi) \Theta(L/2-r)\hat{\bm{\phi}}\right)P_1(\phi,\textbf{r}|s) - \frac{\partial}{\partial \phi} \int d\textbf{r}' \mathcal{T}(\phi-\varphi')\Theta(L/2-r') P_2(\phi,\textbf{r},\textbf{r}'|s),
		\end{split}
	\end{equation}
    where we have needed to introduce the reduced two-constraint distribution
    \begin{equation}
        P_2(\phi,\textbf{r},\textbf{r}'|s) = N(N-1)\int \prod_{k=3}^{N}d\textbf{r}_k P(\phi,\textbf{r},\textbf{r}',\textbf{r}_k|s).
    \end{equation}
    This shows that $P_1$ is part of a hierarchy; $P_1$ depends on $P_2$, which depends on $P_3$, and so on. At this point we have the same difficulty as with equation (\ref{DrFDTgen}); to know $P_1$ we must know the full $N$-constraint dynamics. The self-consistent one-constraint approximation overcomes this. Here an effective dynamics for the one-constraint distribution is supposed which is \textit{independent} of the two constraint dynamics but we \textit{insist} describes the diffusion exactly. This essentially replaces the full system with an effective one comprising the particle and \textit{only one constraint} whose dynamics are determined self-consistently. 

    The scheme is as follows. We suppose that the exact diffusion kernels $\widetilde{D}_r$ and $\widetilde{D}$ are already known. If they are indeed exact, then they must govern the evolution of $P_1$. Similarly, because they are supposed to already contain the interactions between the particles and constraints, removing the interaction of $P_1$ with $P_2$ in (\ref{SEqnPot1}) should not alter the dynamics, as long as $D_r$ and $\mathcal{D}$ are replaced appropriately by $\widetilde{D}_r(s)$ and $\widetilde{\mathcal{D}}(s)$. This gives the effective equation satisfied by $P_1$,
	\begin{equation}
		\label{SEqnPotSC}
		\begin{split}
			s P_1(s) &=  \widetilde{D}_r(s) \frac{\partial}{\partial \phi}\left(\frac{\partial}{\partial \phi} - \mathcal{T} \Theta - g \right)P_1(s) +\widetilde{\mathcal{D}}(s)\bm{\nabla}\cdot \left(\bm{\nabla} + \frac{1}{r} \mathcal{T}\Theta\hat{\bm{\phi}}\right)P_1(s).
		\end{split}
	\end{equation}
    To make sure this is self-consistent, we must use the solution to the effective equation (\ref{SEqnPotSC}) in (\ref{SEqnPot1}). This allows the diffusion kernels to be found self-consistently including binary interactions between the particle and the constraints. Interactions of the particle with more and more constraints can in principle be included by applying the same scheme at a later stage in the hierarchy. 

    As before, we expand $P_1(s)$ about equilibrium to first order in $g$. The first order solution is then, formally,
	\begin{equation}
		\begin{split}
  \label{P1soln}
			\delta P_1(s) = - g \widetilde{D}_r(s) \int \widetilde{G}_1(s) \frac{1}{s} \frac{\partial P^{(1)}_{\text{eq}}}{\partial \phi},
		\end{split}  
	\end{equation}
	where $\widetilde{G}_1$ is the ``one-constraint self-consistent Green function'', the Green function for equation (\ref{SEqnPotSC}). Here $P^{(1)}_{\text{eq}} \approx \rho $ is the equilibrium one particle distribution which satisfies, 
	\begin{equation}
		\label{Peq1useful}
		\frac{\partial P^{(1)}_{\text{eq}}}{\partial \phi} = \mathcal{T} \Theta P^{(1)}_{\text{eq}}, \ \ \ \text{and} \ \ \  \frac{\partial P^{(1)}_{\text{eq}}}{\partial \varphi} = -\mathcal{T} \Theta P^{(1)}_{\text{eq}}.
	\end{equation}
	Using the solution (\ref{P1soln}) in (\ref{FDT3SC}), and following the same manipulations as before, yields a self consistent equation for the diffusion kernel 
	\begin{equation}
		\label{SCDrFDT}
		\frac{1}{\widetilde{D}_r(s)} = \frac{1}{D_r}  + \rho \int \int  \mathcal{T} \Theta \widetilde{G}_1^{\dagger}(s) \mathcal{T} \Theta \equiv \frac{1}{D_r}  + \rho \widetilde{\Delta}(s),
	\end{equation}
	where we have defined the self-consistent one-constraint correction $\widetilde{\Delta}(s)$. This equation is the same result obtained by Szamel and Schweizer \cite{Szamel1993ReptationPolymers, Szamel1994ReptationPolymers} although our derivation is slightly different and connects it to the result of Edwards and Vilgis \cite{Edwards1986TheTransition}.
 
    It must be noted that, since the equation for $\widetilde{G}_1$ also involves the kernel $\widetilde{\mathcal{D}}$, we must in principle supplement (\ref{SCDrFDT}) by another similar equation for $\widetilde{\mathcal{D}}$, the derivation for which is identical. This means we have two coupled equations for $\widetilde{D}_r$ and $\widetilde{\mathcal{D}}$, which may be quite difficult to analyse. For this reason, in this paper, we shall make an assumption for the dynamics of the constraints, expressed through $\widetilde{\mathcal{D}}$. Specifically we suppose that the surrounding constraints have their dynamics slowed in the same way as the particle. This means we can take
    \begin{equation}
        \label{diffassump}
        4 \widetilde{\mathcal{D}}(s) = L^2 \widetilde{D}_r(s).
    \end{equation}
    We argue that this has relevance to the fully three dimensional case because it allows the constraining points to be understood as those points where legs of other particles intersect the plane of our test particle.  However, the full problem should be studied in greater detail in the future.  
	\subsection{Summary of Reptation Results}
	\label{sec:LongDiffRods}
	So that we can compare our forthcoming results for a bent wire frame to rods, here we summarise the behaviour of the angular mean squared displacement for rods predicted by reptation arguments. The arguments are well known in three dimensions \cite{Doi1975RotationalSolution, Doi1978Dynamics1}, and follow straightforwardly in two. 
		
	The reptation picture is based on the supposition that the surrounding constraints effectively force the rod to move within a channel parallel to its length. In two dimensions the width of the channel is $a \sim (\rho L)^{-1}$. The dynamics of the particle are understood by considering how it disengages from the channel. For this to happen, the particle must have diffused sufficiently far, about halfway, along its length. This takes an average time of $\tau_{0} \sim L^2/D_{\parallel} \sim L^2 / D_r$ to occur. Here $D_{\parallel}$  is the longitudinal diffusion coefficient of the rod. Once this has happened, the particle can rotate by an angle $\sim a/L$, after which it is confined to a new channel and so the process repeats. Hence, over sufficiently long times, $t \gg \tau_{0}$, the rotation is visualised as a random walk with step length $a$ and step rate $\tau_{0}$, hence
	\begin{equation}
		\widetilde{D}_{r} \sim \frac{D_{r}}{(\rho L^2)^2}.
	\end{equation}
	We can use this intuition to build up a picture of the angular mean squared displacement, $\mathcal{M}(t)$. For very short times, $t \ll \tau_{0}$, the particle is within the channel and encounters no constraints, hence the particle diffuses as it would in free suspension. For intermediate times, $t \sim \tau_{0}$, the constraints begin to effect the motion and there is some cross over until $t \gg \tau_0$ where the motion is diffusive with the reduced diffusion constant $\widetilde{D}_r$. Therefore the angular mean squared displacement should behave like, 
	\begin{equation}
		\mathcal{M}(t) = \begin{cases} D_r t  \ \ \text{for} \ \ t \ll \frac{L^2}{D_{r}} \\ 
			D_{r} (\rho L^2)^{-2} t \ \ \text{for} \ \ t \gg \frac{L^2}{D_{r}} \end{cases}.
	\end{equation}
	We are interested in how different the result for wire frames will be from this expectation, in the long time limit in particular. 
	
	\subsection{Rotational Diffusion of Wire Frames}
	\label{sec:RotDiffCross}
	Now we consider rotational diffusion of the bent wire frame. The new rotational diffusion kernel is found by solving for $\widetilde{G}_1^{\dagger}(s)$. This satisfies the equation
	\begin{equation}
		\label{CrossTPSCSEqn}
		\begin{split}
			\left(s  - \widetilde{D}_r(s) \frac{\partial^2}{\partial \phi^2} - \widetilde{\mathcal{D}}(s) \nabla^2\right) \widetilde{G}^{\dagger}_1(s)
			-\mathcal{T}(\phi - \varphi) \Theta(L/2 - r) \left(\widetilde{D}_r(s) \frac{\partial}{\partial \phi} - \frac{\widetilde{\mathcal{D}}(s)}{r^2} \frac{\partial}{\partial \varphi}\right)\widetilde{G}^{\dagger}_1(s)&\\
			= \delta(\phi-\phi')\delta(\textbf{r}-\textbf{r}')&,
		\end{split}
	\end{equation}
	where $\mathcal{T}(\phi - \varphi)$ is as given in (\ref{Tdefns1}) and (\ref{Tdefns}). That term supplies the boundary conditions 
 \begin{subequations}
 \label{Cross2PGFBCs}
  	\begin{equation}
		\label{Cross2PGFBCs1}
		\left(\widetilde{D}_r(s) \frac{\partial}{\partial \phi} - \frac{\widetilde{\mathcal{D}}(s)}{r^2} \frac{\partial}{\partial \varphi}\right)\widetilde{G}^{\dagger}_1(s)\Bigg\lvert_{\varphi = \phi}= \left(\widetilde{D}_r(s) \frac{\partial}{\partial \phi} - \frac{\widetilde{\mathcal{D}}(s)}{r^2} \frac{\partial}{\partial \varphi}\right)\widetilde{G}^{\dagger}_1(s)\Bigg\lvert_{\varphi = \phi + \chi} =0,
	\end{equation}
 and
 	\begin{equation}
		\label{Cross2PGFBCs2}
		\left(\widetilde{D}_r(s) \frac{\partial}{\partial \phi} - \frac{\widetilde{\mathcal{D}}(s)}{r^2} \frac{\partial}{\partial \varphi}\right)\widetilde{G}^{\dagger}_1(s)\Bigg\lvert_{\varphi = \phi+\chi}= \left(\widetilde{D}_r(s) \frac{\partial}{\partial \phi} - \frac{\widetilde{\mathcal{D}}(s)}{r^2} \frac{\partial}{\partial \varphi}\right)\widetilde{G}^{\dagger}_1(s)\Bigg\lvert_{\varphi = \phi + 2\pi} = 0,
	\end{equation}
 \end{subequations}
	both are to be satisfied when $r < L/2$. Note that we need two sets of boundary conditions; the first for when the constraint is in the narrower space between the wire frame legs, $\phi< \varphi < \phi + \chi$, and the second for when it is in the wider, $\phi + \chi < \varphi< \phi + 2\pi$. 
	
	We can find an exact solution for $\widetilde{G}_1^{\dagger}$, which is given in appendix \ref{app:CrossTPGF}, and substituting this into (\ref{SCDrFDT}) leads to a closed form for the one-constraint correction 
	\begin{equation}
		\label{Delta1crosses}
		\widetilde{\Delta}(s) = \frac{L^4}{2^4 \widetilde{\mathcal{D}}(s)} \sum_{n \in \text{even}} \sum_{m=1}^{\infty} C_{n}(m) \left(j_{a_n}^2(m) + a_n^2\frac{L^2 \widetilde{D}_r(s)}{ \widetilde{\mathcal{D}}(s)}+\frac{L^2 s}{4 \widetilde{\mathcal{D}}(s)}\right)^{-1}.
	\end{equation}
	where $C_n(m)$ are positive constants which decrease quickly with both $n$ and $m$, $a_n$ are constants depending on $n$ and the opening angle $\chi$ and $j_{a_n}(m)$ is the $m^{\text{th}}$ zero of the Bessel function of order $a_{n}$. All concrete definitions may be found in appendix \ref{app:CrossTPGF}.
	
	To extract $\widetilde{D}_r$, we need to use the assumption (\ref{diffassump}) for $\widetilde{\mathcal{D}}$. Even with this, the form of $\widetilde{\Delta}$ is a quite unwieldy. However, this can be resolved by assuming that the sum in (\ref{Delta1crosses}) is dominated by its first term, which is reasonable due to the decreasing behaviour of $C_n(m)$. Under these assumptions we may write the one-constraint correction as,
	\begin{equation}
		\label{Delta1Approx}
		\widetilde{\Delta}_1(s) = \frac{a L^2}{b \widetilde{D}_r(s)+ s},
	\end{equation}
	where $a$ and $b$ are positive constants whose values are qualitatively irrelevant. It then follows that the new diffusion kernel is 
	\begin{equation}
		\label{fullDr}
		\frac{\widetilde{D}_r(s)}{D_r} = \frac{1}{2} \left(\left(1 - c/c^{*}\right) - \tau_0 s\right) + \frac{1}{2} \sqrt{\left(\left(1 - c/c^{*}\right) - \tau_0 s\right)^2 + 4 \tau_0 s}.
	\end{equation}
	where $c = \rho L^2$, $c^{*} = b/a$ and $\tau_0 = D_r^{-1}$. 
	This can be used to compute the angular mean squared displacement, $\mathcal{M}(s)$ from (\ref{MSDDr}). Unfortunately, we cannot find a simple closed form for $\mathcal{M}(t)$, although it may be written in terms of an integral involving a Bessel function (see Appendix \ref{app:CrossMSD}). Nonetheless, we are able to compute the long and short time limits easily.
	
	The short time limit has $t \ll \tau_0$, so we take $\tau_0 s \gg 1$ in (\ref{fullDr}), this yields,
	\begin{equation}
		\mathcal{M}(s) \sim \frac{2 D_r}{s^2}. 
	\end{equation}
	After taking the inverse Laplace transform we get, 
	\begin{equation}
		\mathcal{M}(t) \sim 2 D_r t,
	\end{equation}
	so that the short time motion is diffusive with diffusion constant $D_r$, as in free suspension. This is to be expected since for short times $t \ll \tau_0$, the particle has not diffused far enough to begin to interact with the constraints. 
	
\begin{figure}\includegraphics[width=16cm]{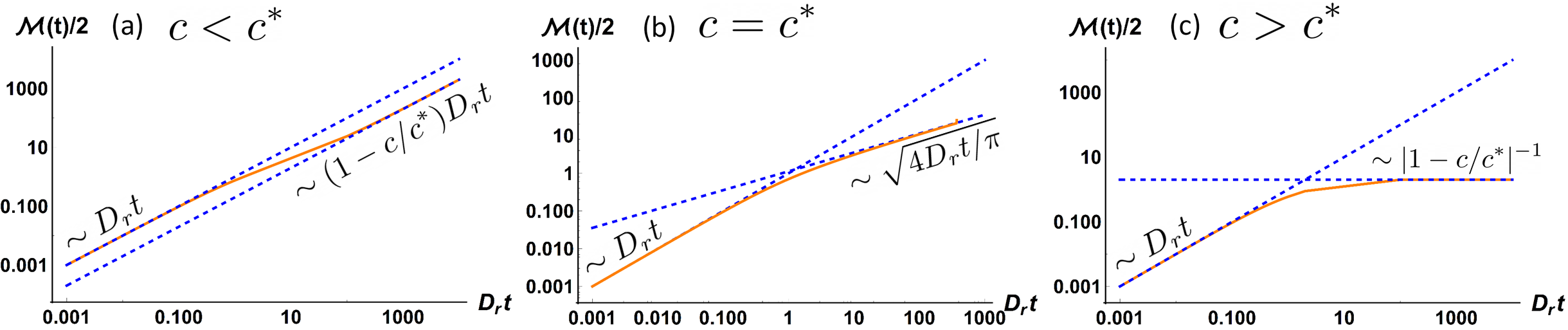}
		\caption{\label{fig:MSDCross} Log-log plots of the normalised angular mean squared displacement $\mathcal{M}/2$ as a function of reduced time $D_r t$ for three different concentrations $c$. In each case the orange curve is the numerical result computed from (\ref{fullDr}) and the two dashed lines are the long and short time asymptotics. Panel (a) has $c < c^*$ (specifically $c = 0.8 c^*$), and the behaviour clearly remains diffusive. Panel (b) has $c = c^*$, and the long time sub-diffusive $\sim t^{1/2}$ scaling is seen. Panel (c) has $c>c^*$ (specifically $c = 1.5 c^*$) and $\mathcal{M}(t)$ clearly plateaus.}
	\end{figure}

	For long times, $t \gg \tau_0$, the behaviour is more complicated, depending on the ratio $c/c^{*}$. The behaviour of $\mathcal{M}_r(s)$ for $s \tau_0 \ll 1$ is summarised as follows,
	\begin{equation}
		\mathcal{M}(s) \sim \begin{cases} \frac{2 (1-c/c^{*})}{\tau_0 s^2} \ \ &\text{if} \ \ c/c^{*} < 1\\ 
			\sqrt{4 \frac{D_r}{s^3}} \ \ &\text{if} \ \ c/c^{*} = 1\\ 
			\frac{2}{|1-c/c^{*}| s} \ \ &\text{if} \ \ c/c^{*} > 1 \end{cases}.
	\end{equation}
	This leads to long time behaviour which is very different from the reptation result. Inverting the Laplace transform we find,
	\begin{equation}
		\label{MSDcrosses}
		\mathcal{M}(t) \sim \begin{cases} 2 (1-c/c^{*})D_r t \ \ &\text{if} \ \ c/c^{*} < 1\\ 
			\frac{4}{\sqrt{\pi}}\left(D_r t\right)^{1/2} \ \ &\text{if} \ \ c/c^{*} = 1\\ 
			\frac{2}{|1-c/c^{*}|} \ \ &\text{if} \ \ c/c^{*} > 1 \end{cases}.
	\end{equation}
	These asymptotics are plotted against $\mathcal{M}(t)$ computed from (\ref{fullDr}) and (\ref{MSDDr}) in Fig.(\ref{fig:MSDCross}) for three concentrations; panel (a) has $c < c^*$, panel (b) has $c = c^*$, and panel (c) has $c > c^*$. This shows that there is a critical concentration, $c^{*}$, at which the dynamics radically change. The approximate analysis presented here gives a value $c^* \approx 16$, although we do not expect this to be quantitatively accurate. Below the critical concentration, the motion is diffusive, with a diffusion constant which is linearly reduced in the density $D_r \to D_r (1 - c/c^{*})$. Precisely at the critical concentration sub-diffusive behaviour is predicted, with the angular mean squared displacement increasing $\sim \sqrt{t}$. Above this critical concentration, all motion stops after some time with $\mathcal{M}(t)$, approaching a constant. We note that this constant becomes very small for large concentrations $c/c^{*} \gg 1$ and scales in same way as the average angle the particle would need to rotate by before it encounters a constraint $\sim c^{-1}$. This is consistent with the qualitative picture presented in \cite{King2021ElasticModelc} where, after a short equilibration time $\tau_0$, the particle is effectively confined to a cage of angular size $\sim (\rho L^2)^{-1}$.
	
	This behaviour is entirely different from that of rods and indicates how strong a role the topological constraints have for wire frame particles. However, it is entirely possible that these results may be changed when the interactions with more constraints are included. On the other hand, it is important to mention that the approximating $\widetilde{\Delta}_1$ by only the first term in the sum (\ref{Delta1crosses}) does not severely effect the behaviour of $\mathcal{M}(t)$, changing only numerical factors in (\ref{MSDcrosses}) not the scaling with $t$. In any case, this shows that the dynamics of these particle shapes can be quite complicated in the absence of reptation. 
	
	\section{Rheology}
	\label{sec:Rheology}	
	In the previous section we have demonstrated that the constrained dynamics of cross shaped wire frames are very different from those of rods. Here we consider how these differences affect the relaxation of the stress in the linear regime where the applied strain is small. 
    We would also like to investigate the magnitude of the elastic stress. Previous work has shown that the initial value of the elastic stress is approximately independent of concentration for rods, but increases as \textit{the square} of the concentration for wire frames \cite{King2021ElasticModelc} and our results should be consistent with this.
	\subsection{Derivation of the Stress Tensor}
	When we discuss the rheology of this model, we will consider imposing a particular solvent fluid flow described by the strain rate tensor $\kappa_{\alpha \beta}$. The goal is then to determine the resulting stress tensor $\sigma_{\alpha \beta}$ to linear order in $\kappa$ and how this is modified away from the Newtonian result by the presence of the particle, constraints and their interactions. To keep our discussion as simple as possible, and to connect this work to that done previously \cite{King2021ElasticModelc,King2021ElasticSystemsc}, we restrict our attention exclusively to simple shear flow where $\kappa$ has only one non-vanishing component, $\kappa_{xy}=\dot{\gamma}$. In this case, $\sigma_{\alpha \beta}$ will also have only one component, the $xy-$component, which we shall refer to as $\sigma$. Note that the pressure may be dropped because we assume the solvent is incompressible.   
	
	It is useful to divide $\sigma$ into three parts, 
	\begin{equation}
		\sigma =  \eta_{s} \dot{\gamma} + \sigma^{e} + \sigma^{v}.
	\end{equation}
	The first part is the Newtonian contribution from the solvent, which is irrelevant for our comparison of the rheology of the rod and cross systems; only the final two parts include the particle contribution to the stress. These are the elastic, $\sigma^{e}$, and viscous, $\sigma^{v}$ parts respectively. To derive expressions for these quantities we need to understand how the particle and constraints move in the presence of the shear. We assume that the constraints follow the shear flow like spherical particles. This means that the velocity of the $i^{\text{th}}$ constraint induced by the shear flow is
 \begin{equation}
      \textbf{v}_i = \dot{\gamma} \left(r_i \sin \varphi_i \cos \varphi_i \ \hat{\textbf{r}}_i - r_i \sin^2\varphi_i \ \hat{\bm{\varphi}}_i\right).
 \end{equation}
 This is written in polar co-ordinates with radial and angular unit vectors $\hat{\textbf{r}}_i$ and $\hat{\bm{\varphi}}_i$ respectively. 
 It is a standard hydrodynamic problem to deduce the angular velocity of the bent wire frame induced by the shear flow (details may be found in \cite{Makino2004ViscoelasticityShape,King2020ParticleSuspensionsb,King2021EffectsDynamics}) the result is
     \begin{equation}
     \label{angvels}
         \omega = -\frac{1}{2}\dot{\gamma} \left(\sin^2\phi + \sin^2(\phi+\chi)\right)  
     \end{equation}
	
With these expressions, there are many ways for us to derive the elastic and viscous stresses. Here we will discuss their derivations separately, since it allows us to highlight a difficulty with the viscous part.
	\subsubsection{Elastic Stress}
	We derive the elastic stress from the virtual work principle relating the variation of the free energy under deformation to the stress. In this section we shall only outline the derivation, more details may be found in Appendix \ref{app:crossstress}. 
	
	Suppose that the system is infinitesimally deformed by a strain $\gamma$. This changes the free energy by $\delta F$. This change in free energy is related to the elastic stress by
	\begin{equation}
		\label{virtualwork}
		\delta F = \sigma^{e} \gamma.
	\end{equation}
	We find the elastic stress by determining $\delta F$ and comparing to this form. 
	
	The free energy is written, using the short hand (\ref{Shorthand}), as
	\begin{equation}
		\label{FenShort}
		F = k_B T\int P \log \frac{P}{P_{\text{eq}}}.
	\end{equation}
	The inclusion of the equilibrium distribution $P_{\text{eq}}$ allows us to take into account the collision forces between the particle and the constraints by keeping track of its derivatives (\ref{Pequseful}). From (\ref{FenShort}) we can directly compute its infinitesimal change in terms of that of the probability distribution, $\delta P$,
	\begin{equation}
		\label{deltaF}
		\delta F = k_B T\int \delta P \left(\log \frac{P}{P_{\text{eq}}} + 1\right).
	\end{equation}
	We compute $\delta P $ from the Smoluchowski equation. In the presence of the shear flow, (\ref{SEqn}) is modified to
	\begin{equation}
		\label{SEqnCrossK}
		\begin{split}
			\frac{\partial P}{\partial t} &=  D_r \frac{\partial}{\partial \phi}\left(\frac{\partial}{\partial \phi} - \sum_{i=1}^{N} \mathcal{T}_i \Theta_i \right)P + \sum_{i=1}^{N} \mathcal{D}\bm{\nabla}_i \cdot \left(\bm{\nabla}_i + \frac{1}{r_i} \mathcal{T}_i \Theta_i \hat{\bm{\phi}}\right)P \\
			& - \dot{\gamma} \sum_{i=1}^{N} \left(r_i \sin \varphi_i \cos \varphi_i \frac{\partial P}{\partial r_i} + \sin^2 \varphi_i \frac{\partial P}{\partial \varphi_i}\right) + \frac{\dot{\gamma}}{2} \frac{\partial}{\partial \phi}\bigg(\left(\sin^2\phi + \sin^2(\phi+\chi)\right) P \bigg).
		\end{split}
	\end{equation}
 During the course of the deformation, the strain rate $\dot{\gamma}= \gamma/\delta t$ dominates the evolution so we find from (\ref{SEqnCrossK}) that,
	\begin{equation}
		\label{CrossDeltaP}
		\delta P = - \dot{\gamma} \sum_{i=1}^{N} \left(r_i \sin \varphi_i \cos \varphi_i \frac{\partial P}{\partial r_i} + \sin^2 \varphi_i \frac{\partial P}{\partial \varphi_i}\right) + \frac{\dot{\gamma}}{2} \frac{\partial}{\partial \phi}\bigg(\left(\sin^2\phi + \sin^2(\phi+\chi)\right) P \bigg).
	\end{equation}
	This is substituted into (\ref{deltaF}) and integrated by parts twice to give,
	\begin{equation}
		\label{StressCross}
  \begin{split}
 		\sigma^{e} &= k_B T \cos \chi \int P \sin(2\phi +\chi)\\
   &+ \frac{1}{2} k_B T \int P \sum_{i=1}^{N} \left(\sin^2\phi + \sin^2(\phi+\chi) -2\sin^2\varphi_{i} \right) \mathcal{T}(\phi - \varphi_i) \Theta(L/2 - r_i).     
  \end{split}
	\end{equation}
     We would obtain the same result for the elastic stress if we had used the standard Kirkwood approach \cite{Doi1986TheDynamics,Kirkwood1949TheBehavior,Kirkwood1954TheMacromolecules}. This expression has two parts. The first, 
	\begin{equation}
		\label{IntStressCross}
      		\sigma^{e}_{\text{Int}}(t) = k_B T \cos \chi \int P(t) \sin(2\phi +\chi) \equiv k_B T \left\langle S_{\text{Int}}(t)\right\rangle,
	\end{equation}    
    is the ``internal'' or``\textit{intra}-molecular'' stress whose origin is entropic and depends on the orientational relaxation of the particle. The second,
    \begin{equation}
		\label{ExtStressCross}
  \begin{split}
		\sigma^{e}_{\text{Ext}}(t) &= \frac{1}{2} k_B T \int P(t) \sum_{i=1}^{N} \left(\sin^2\phi + \sin^2(\phi+\chi) -2\sin^2\varphi_{i} \right) \mathcal{T}(\phi - \varphi_i) \Theta(L/2 - r_i) \\
  &\equiv k_B T \left\langle S_{\text{Ext}}(t)\right\rangle.
  \end{split}
	\end{equation}
    is the ``external'' or ``\textit{inter}-molecular'' stress originating from the forces between the particle and constraints. For convenience, on the right hand side of both of these equations, we have defined the quantities $S_{\text{Int}}$ and $S_{\text{Ext}}$ which are averaged over $P(t)$ to give the internal and external stresses respectively. 
    
    Notice from (\ref{ExtStressCross}) that as the bent wire frame approaches a rod, $\chi \to \pi$, the external stress vanishes. On the other hand, from (\ref{IntStressCross}), as it approaches an L-shape, $\chi \to \pi/2$, the internal stress vanishes. The implication of this that when $\chi = \pi/2$ the elastic stress is \textit{zero identically} if no constraints are present. This result is re-assuring since, for fundamental symmetry reasons, the stress originating from a single L-shaped particle in a Newtonian fluid must vanish \cite{King2020ParticleSuspensionsb}. 

    The ``internal/external'' nomenclature is due to Fixman, who studied these two contributions in simulations of flexible polymer chains with fixed joint angles \cite{Fixman1991StressSolutions}. In that case he found that the external stress dominated over the internal stress. This points to the essential difference between the stress in wire frame and rod suspensions and why the the stress in the former is significantly larger.
	
	\subsubsection{Viscous Stress}
	The viscous response is determined from the hydrodynamic energy dissipation under the prescribed velocity gradient. This is affected not only by the kinetic constraints but also by the hydrodynamic interactions between the particle and constraints. This is because of rapid, small scale motions which dissipate energy but for which the kinetic constraints are not relevant. We ignore hydrodynamic interactions in our model thus preventing us from treating the viscous stresses accurately. The hydrodynamic interactions may be taken into account with a self-consistent effective medium approach \cite{Muthukumar1983ScreeningMacromolecules, Edwards1974TheorySolutions, Freed1974PolymerSolutions} or by using sophisticated diagrammatic techniques which allow the interactions between an infinite number of particles to be included \cite{Fredrickson1989HeatFibers, Shaqfeh1990TheRods}. These approaches essentially result in the determination of an effective rotational friction coefficient, which can be used to find the viscous stress by single particle arguments. Note that there is no theoretical reason for the new friction constant to be related to $\widetilde{D}_r$ \cite{Doi1986TheDynamics}. These calculations are challenging, so we shall neglect the viscous stress by restricting our attention to a specific, illustrative case; step strain. In this case, the applied strain rate is \textit{always} given by
	\begin{equation}
		\dot{\gamma}(t) = \gamma \delta(t).
	\end{equation}
	Hence, the viscous stress is also proportional to a delta function which allows us to ignore it since it vanishes on all practical timescales. 
	
	It is worth noting that while neglecting hydrodynamic interactions impacts the viscous stress, it does not effect $\widetilde{D}_r$ or the elastic stress so severely. It has been shown previously in concentrated rod systems that, because they concern the long time relaxation of the system to equilibrium, these properties are significantly more sensitive to the entanglements than they are to hydrodynamic interactions \cite{Muthukumar1983ScreeningMacromolecules,Doi1986TheDynamics}. 
	\subsection{Summary of Repation Results}
	Let us summarise the well known results for rods from reptation theory \cite{Doi1978Dynamics1,Doi1978Dynamics2}. We shall compare these to our results for wire frames in the next section. 
	
	The elastic stress for a single rod in two dimensions in the reptation theory is given by (\ref{IntStressCross}) with $\chi = \pi$, and we ignore the viscous stress for the reasons discussed previously. For the elastic stress, it is necessary to solve for the orientational distribution to linear order in the shear rate. The reptation theory tells us that for short times, $t \ll D_r^{-1} (\rho L^2)^{-2}$, the motion of the particle is un-hindered so the orientation relaxes as in free suspension, and the linear elastic stress in this regime is,
	\begin{equation}
 \label{shorttimerep}
		\sigma^{e} = \frac{1}{2}k_B T e^{- 4 D_r t} \ \ \ \text{for} \ \ \ t \ll D_r^{-1} (\rho L^2)^{-2}.
	\end{equation}
	For longer times, when the entanglement effect takes over, the motion of the particle is governed by the modified diffusion constant $\widetilde{D}_r \sim D_r (\rho L^2)^{-2}$ given by the reptation theory \cite{Doi1975RotationalSolution,Doi1986TheDynamics}. Hence, in this limit, the elastic stress is 
	\begin{equation}
		\sigma^{e} \sim \frac{1}{2}k_B T e^{- 4 D_r t /(\rho L^2)^2} + C\ \ \ \text{for} \ \ \ t \gg D_r^{-1} (\rho L^2)^{-2},
	\end{equation}
	where $C$ is a constant which is needed for the stress response to be continuous. 
	
	We expect that the general structure of the stress should always be of this form, even for wire frames. For short times, when the particles have not diffused far enough to interact with their surroundings, the elastic stress should be exactly as found in dilute suspensions. Then there is a crossover to the long time behaviour where the decay of the elastic stress is much slower and is controlled by the interactions of the particle with its surroundings. The particular form of the long time decay, however, is where we expect to see differences. 
	\subsection{Wire Frame Elastic Stress}
	\label{sec:StressCross}
    \subsubsection{Relaxation Modulus}
    In the linear regime, where the step strain $\gamma$ is sufficiently small, the elastic stress can be written in terms of the relaxation modulus, $\Gamma(t)$,
    \begin{equation}
        \sigma^{e}(t) = \int_{-\infty}^{t} dt' \ \Gamma(t-t') \dot{\gamma}(t') = \gamma \Gamma(t).
    \end{equation}
    The relaxtation modulus describes the magnitude and decay of the stress. The goal of this section is to understand its short and long time structure.  We begin with its expression as the following equilibrium average \cite{Doi1986TheDynamics,Fixman1991StressSolutions},
    \begin{equation}
        \Gamma(t) = k_B T \left\langle \left(S_{\text{Int}}(t)+S_{\text{Ext}}(t)\right)\left(S_{\text{Int}}(0)+S_{\text{Ext}}(0)\right) \right\rangle_{\text{eq}}.
    \end{equation}
    Hence it may be split into three parts, 
    \begin{subequations}
        \begin{equation}
        \Gamma(t) = \Gamma_{\text{Int}}(t) + \Gamma_{\text{Ext}}(t) + \Gamma_{\text{IE}}(t).
    \end{equation}
    The first is the solely internal contribution 
    \begin{equation}
    \label{GammaInt}
        \Gamma_{\text{Int}}(t) = k_B T \left\langle S_{\text{Int}}(t) S_{\text{Int}}(0) \right\rangle_{\text{eq}},
    \end{equation}
    the second is entirely external 
    \begin{equation}
    \label{GammaExt}
        \Gamma_{\text{Ext}}(t) = k_B T \left\langle S_{\text{Ext}}(t) S_{\text{Ext}}(0) \right\rangle_{\text{eq}},
    \end{equation}
    and finally the third is the cross term
     \begin{equation}
     \label{GammaIE}
        \Gamma_{\text{IE}}(t) = k_B T \left\langle S_{\text{Int}}(t) S_{\text{Ext}}(0) + S_{\text{Ext}}(t) S_{\text{Int}}(0)\right\rangle_{\text{eq}}.
    \end{equation}
    \end{subequations}
    
    Using the definitions (\ref{IntStressCross}) and (\ref{ExtStressCross}) and some trigonometric identities (see appendix \ref{app:relax} for details), we may manipulate each of these to obtain more useful expressions;
    \begin{subequations}
    \label{fullmod}
           \begin{equation}
    \label{IntMod}
         \Gamma_{\text{Int}}(t) = \frac{1 }{2} k_B T \cos^2 \chi \left\langle \cos\left[2 (\phi(t)-\phi(0)) \right] \right\rangle_{\text{eq}}.
    \end{equation}
        \begin{equation}
    \label{ExtMod}
         \Gamma_{\text{Ext}}(t) = \frac{1 }{2} k_B T \sin^2 \chi \left\langle \cos\left[2 (\phi(t)-\phi(0)) \right] \sum_{i,j=1}^{N} \mathcal{T}^{\dagger}_i(t) \Theta_i(t)\mathcal{T}^{\dagger}_j(0) \Theta_j(0)\right\rangle_{\text{eq}},
    \end{equation}
        \begin{equation}
    \label{MixedMod}
        \Gamma_{IE}(t) = k_B T \sin \chi \cos \chi  \left\langle \cos\left[2 (\phi(t)-\phi(0)) \right] \sum_{i=1}^{N} \left(\mathcal{T}^{\dagger}_i(t)\Theta_i(t) + \mathcal{T}^{\dagger}_i(0)\Theta_i(0) \right)\right\rangle_{\text{eq}}.
    \end{equation}
    \end{subequations}
    Here we have defined $\mathcal{T}^{\dagger}_i = \delta(\varphi_i - \phi) + \delta(\varphi_i - \phi - \chi)$ and the arguments of $\mathcal{T}^{\dagger}_i$ and $\Theta_i$ indicate the time at which their co-ordinate inputs should be evaluated.

    Notice that each of these contains a factor of $\cos[2 (\phi(t)- \phi(0))]$. This is the part that controls the stress relaxation and relates it to the return of the orientation of the particle to equilibrium after the step strain. Each term also comes with a geometric factor; the function of $\chi$ outside the equilibrium average. Notice that when the particle is not bent, $\chi = 0$ or $\pi$, only the \textit{internal} modulus survives and when the particle is L-shaped, $\chi = \pi/2$, only the \textit{external} modulus is left. The external and mixed moduli also come with factors representing the collision forces between the particle and constraints; the sums containing $\mathcal{T}^{\dagger}_i(t) \Theta_i(t)$ which ensure that the $i^{th}$ constraint is in contact with the particle at time $t$. These terms depend on the number of constraints and so control the scaling of the elastic stress with constraint density $\rho$. 
    \subsubsection{Initial Magnitude}
    From equations (\ref{fullmod}) we can estimate both the initial size and time dependence of the relaxation modulus relatively easily. Let us begin with the magnitude after the step strain, which may be deduced essentially from dimensional analysis. When the equilibrium average is taken to compute the relaxation modulus the positions of all $N$ constraints must be integrated over along with the orientation of the particle. The equilibrium distribution gives a factor of $V^{-1}$ for each of the $N$ constraints. Every $\mathcal{T}^{\dagger}_i$ has no dimensions and just fixes the angle $\varphi_i$, but the $\Theta_i$ constraints the radial integral to $0 < r_i < L$. This means that when a $\Theta_i$ appears in the equilibrium average it gives a factor of
    \begin{equation}
        \int_{0}^{L} \frac{dr}{V} \ r \sim \frac{L^2}{V} 
    \end{equation}
   while the result would be unity if it were not present. The internal modulus (\ref{IntMod}) has no $\Theta_i$s appearing and is therefore independent of $\rho$. The mixed modulus (\ref{MixedMod}) has $N$ terms, each containing one $\Theta_i$ and so contributes a factor of $\sim N L^2 / V = \rho L^2$. Finally the external modulus (\ref{ExtMod}), which has $\sim N^2$ terms with two $\Theta_i$s, providing a contribution $\sim N^2 L^4 /V^2 = (\rho L^2)^2$. Hence, for sufficiently large density this is the dominant term and the relaxation modulus would be approximately
   \begin{equation}
       \Gamma(t) \sim (\rho L^2)^2 k_B T \sin^2 \chi \left\langle \cos\left[2 (\phi(t)-\phi(0)) \right] \right\rangle_{\text{eq}}.
   \end{equation}
Note that, in writing the relaxation modulus in this form, we have assumed that the averages over the constraints may be taken separately from the average over the orientation of the particle. This is justified within the one-constraint self-consistent approximation introduced in section \ref{sec:MSDselfconsapp}, which essentially replaces the interactions between the particle and constraints by modifying the rotational diffusion kernel of the particle. Also notice that this is precisely the same scaling with $\rho$ found using a geometric method which related the elastic stress to the change under shear deformation of the size of the region to which the particle is constrained \cite{King2021ElasticModelc,King2021ElasticSystemsc}. For low concentrations, however, it is the internal modulus which dominates. The ratio of these two contributions depending on $\chi$ is also the same as the geometric theory; $\Gamma_{\text{Ext}}/\Gamma_{\text{Int}} \sim (\rho L^2)^2 \tan^2 \chi$. We are then led to the same conclusion that the magnitude of the elastic stress is extremely sensitive to particle geometry in these systems, with wire frames which are bent through an angle of more than $\chi_c \sim (\rho L^2)^{-1}$ having significantly larger elasticity than straight rods.  
\subsubsection{Time Dependence} 
The time dependence of the relaxation modulus is determined by 
\begin{equation}
    R(t)=\left\langle \cos[2(\phi(t)-\phi(0))]\right\rangle_{\text{eq}}.
\end{equation}
One might expect that this would require another calculation along the lines of that presented in section \ref{sec:RotDiffCross}, but we can re-use our calculation of the mean squared displacement to avoid this extra effort. The first step is to write the cosine as the real part of a complex exponential which is then expanded as a series
\begin{equation}
    R(t) = \Re \left\langle e^{ 2 i \left(\phi(t)-\phi(0)\right)}\right\rangle_{\text{eq}} = \Re \sum_{n=0}^{\infty} \frac{(2i)^n}{n!}\left\langle \left(\phi(t)-\phi(0)\right)^n \right\rangle_{\text{eq}}.
\end{equation}
This reduces the problem to finding the moments,
\begin{equation}
    \mathcal{M}_n(t) = \left\langle \left(\phi(t)-\phi(0)\right)^n \right\rangle_{\text{eq}}.
\end{equation}
We argue that, at the level of the self-consistent one-constraint approximation, this should be done assuming that the interactions between the particle and constraints are all accounted for by the modified diffusion kernels $\widetilde{D}_r$ and $\widetilde{\mathcal{D}}$. It is then easy to show (see appendix \ref{app:rec}) that the moments satisfy the recursion relation
\begin{equation}
\label{Mnrec}
    \mathcal{M}_n(s) = n(n-1) \frac{\widetilde{D}_r(s)}{s} \mathcal{M}_{n-2}(s),
\end{equation}
which we have written in Laplace transform. This two step recursion relation has the initial conditions,
    \begin{equation}
    \label{Mnrecint}
        \mathcal{M}_1(s) = 0, \ \ \ \text{and} \ \ \ \mathcal{M}_2(s) = \frac{\widetilde{D}_r(s)}{s^2}.
    \end{equation}
The first of these makes all the odd moments vanish and the second relates all even moments to the angular mean squared displacement (\ref{MSDDr}). We can now solve (\ref{Mnrec}) for the non-vanishing even moments
\begin{equation}
    \mathcal{M}_{2n}(s) = \frac{(2n)! \widetilde{D}_r^n(s)}{s^{n+1}}. 
\end{equation}
Hence we have
\begin{equation}
\label{Rs}
    R(s) = \frac{1}{s}\sum_{n=0}^{\infty} (-1)^n \left(\frac{4 \widetilde{D}_r(s)}{s}\right)^n.
\end{equation}
The time dependence of the elastic stress can therefore be understood by taking the inverse Laplace transform of the above using equation (\ref{fullDr}) for $\widetilde{D}_r$. It is tempting take the sum because it is a geometric series. Unfortunately, the need for the inverse Laplace transform ruins this possibility since, for most of the required $s$ integral, $| 4 \widetilde{D}_r / s | > 1$ and the series diverges. Nevertheless, the short- and long-time limits can be determined straightforwardly by refering to the results in section \ref{sec:RotDiffCross} (for additional details see appendix \ref{app:time}).

For short times such that $t \ll \tau_0$, we have $s \tau_0 \gg 1$. This results in, 
\begin{equation}
\label{Rtshort}
    R(t) \sim e^{- 4 D_r t}, \ \ \ \text{for} \ \ t \ll \tau_0.
\end{equation}
This is consistent with our expectations when comparing to the reptation theory (\ref{shorttimerep}); before the particle has diffused far enough to interact with the constraints, the stress relaxes just as in dilute suspension. 

The long time limit, with $t \gg \tau_0$ or $s \tau_0 \ll 1$, is more complicated depending on $\rho$. Once again the behaviour is determined by the proximity of $c = \rho L^2$ to $c^*$ in (\ref{fullDr}). The analysis in this case is a little more involved (see appendix \ref{app:time}). The results are
\begin{equation}
\label{rt}
		R(t) \sim \begin{cases} e^{- 4 D_r(1 - c/c^{*})t} \ \ &\text{if} \ \ c/c^{*} < 1\\ 
			\frac{1}{4\sqrt{ \pi D_r t}} \ \ &\text{if} \ \ c/c^{*} = 1\\ 
			\frac{c/c^{*}-1}{ c/c^{*}+ 3} \ \ &\text{if} \ \ c/c^{*} > 1 \end{cases}.
	\end{equation}
From equation (\ref{fullmod}) we see that $R(t)$ controls the time dependence of the stress. Hence (\ref{rt}) shows that, much like the mean squared displacement, the relaxation of the stress for wire frame particles is radically different from rods. The reptation theory shows that the elastic stress always decays exponentially for long times no matter the concentration, albeit with an extremely long decay time. For wire frames, on the other hand, the stress can decay either exponentially, algebraically $\sim t^{-1/2}$ or eventually plateau and persist indefinitely, depending on the concentration. It should be noted that the self-consistent one-constraint approximation used to derive this result cannot take into account the co-operative motions between large numbers of particles. These motions would allow the particles to slowly re-arrange over very long times and lead to a stress which eventually decays. 
	\section{Discussion and Conclusions}
	\label{sec:Conclusions}
	In this paper we studied the dynamics of wire frame particles in concentrated suspension by means of a simplified two dimensional model. Here, a single particle is placed in the plane and is surrounded by point like constraints which the particle cannot cross. The dynamics are included by allowing the particle and the constraints to diffuse. For simplicity we only considered one wire frame shape; a bent particle constructed from two equal length rods which meet at an angle $\chi$. We were interested in comparing the behaviour of the wire frame to that of rods. 
	
	The dynamics of the particle were studied via the angular mean squared displacement. The centre of the particle was assumed to be fixed, since the translational diffusion is extremely slow in these systems \cite{VanKetel2005StructuralGas,Sussman2011MicroscopicMacromolecules}. The mean squared displacement was written in terms of a general time dependent diffusion kernel, which we obtained using a self-consistent approximation to include binary interactions between the particle and constraints. This same approximation has been used previously \cite{Szamel1993ReptationPolymers, Szamel1994ReptationPolymers} to study the reptation of rod-like particles.
	
	The rotational diffusion of bent wire frames was found to be markedly different from that of rods. At high concentrations $\rho L^2  \gg 1$ rods are able to diffuse via reptation, although their rotational diffusion constant is strongly suppressed \cite{Doi1975RotationalSolution, Doi1978Dynamics1}. The mean squared displacement of wire frames, on the other hand, has a rich structure depending on the concentration. For short times, no matter the concentration, the diffusion is the same as in free suspension. For longer times, there is a critical concentration at which the mean squared displacement transitions from diffusive, with a diffusion constant linearly reduced in the concentration, to sub-diffusive, increasing $\sim t^{1/2}$. Above this critical concentration, we find a complete cessation of motion, with the angular mean-squared displacement plateauing at a value equivalent to the average angle between two constraints. This indicates that the particle becomes confined to a small cage for long periods of time. 
	
	We also studied the rheology in this model, specifically the elastic stress in response to step shear strain. We were interested in two aspects in particular; the scaling of its initial value with constraint concentration and its subsequent decay. We find that the elastic stress for rods is purely \textit{intra}-molecular, and so its initial value must be approximately independent of concentration. For wire frames, even if they are only slightly bent $\chi < 170^{\circ}$ say, the \textit{inter}-molecular stress becomes very important. This leads to an initial value of the stress proportional the constraint concentration squared. This is consistent with previous results, which used a geometric method to estimate the initial elastic stress \cite{King2021ElasticModelc, King2021ElasticSystemsc}.
	
	The self consistent diffusion kernel can then be used to study the relaxation of the stress. This lead to a very different decay of the elastic stress for wire frames than for rods. The reptation theory tells us that, for rods, the elastic stress initially decays exponentially as in free suspension but changes to a slower exponential relaxation for long times. For wire frames, while the short time decay is also exponential as in free suspension, the long time behaviour strongly depends on concentration. Below the critical concentration, the decay is exponential with a modified decay rate. At the critical concentration, the elastic stress decays algebraically $\sim t^{-1/2}$. Above this concentration, the stress appears to persist indefinitely at its very large initial value. This demonstrates an obvious, measurable, qualitative difference between concentrated wire frame and rod suspensions. These differences should be very sensitive to the shape of the particle so may have application to the design of DNA nano-star systems for particular purposes \cite{Xing2018MicrorheologyHydrogels}.
	
	We must note that we neglected the viscous stress by considering only step shear. This meant that the viscous stress was proportional to a delta function, so it could be ignored on practical timescales. We did this because we assumed no hydrodynamic interactions between the constraints and the particles. This is reasonable at low concentrations, but more delicate in concentrated suspension. This does not effect our results for the diffusion kernel or elastic stress because these are most strongly effected by the kinetic constraints. The viscous stress, on the other hand, is sensitive to fast, small scale motions which dissipate energy and are controlled by hydrodynamics but for which the non-crossing constraints are irrelevant. Understanding this contribution to the stress is important and it should be possible to use pre-existing methods to study it \cite{Muthukumar1983ScreeningMacromolecules,Edwards1974TheorySolutions, Freed1974PolymerSolutions, Shaqfeh1990TheRods}, but given how large the elastic stress is in the wire frame systems we expect that to be the dominant contribution. 

	There are a number of ways in which our work may be extended and improved. Clearly we must consider three dimensions. The same machinery laid out here and by Sussman and Schweizer \cite{Sussman2011MicroscopicMacromolecules} can be used, although it may be technically challenging. It should be noted that this two dimensional model has already been shown to produce results which carry over qualitatively to three dimensions \cite{King2021ElasticModelc, King2021ElasticSystemsc}, and we expect that to be the case for the results here. 
	
	It is also necessary to return to some of the simplifications made in this article. For example, we did not include translation rotation coupling when studying the diffusion. Similarly, we made assumptions about the modified diffusion kernels for the constraints, rather than determining them within our framework. If this is done, then it may allow the possibility of studying mixtures of different wire frame particle shapes. A further assumption we made is that there is no spatial dependence to the diffusion kernel. Introducing this may allow us to study non-uniform systems or instabilities. Along the same lines, in only studying the linear rheology with homogenous background flow, we overlook the possibility of flow instabilities. This certainly warrants further investigation given the possibility of non-monatonic stress-strain relationships in wire frame suspensions \cite{King2021ElasticModelc,King2021ElasticSystemsc}.
	
	By far the most important simplification we have made is only including binary interactions between the particle and constraints. Including multi-particle interactions beyond this will likely need an alternative treatment. There are at least two possibilities. The first is to develop equation (\ref{DrFDTgen}) for $\widetilde{D}_r$ as a formal series in powers of $\rho$. Then, one particular co-operative motion may be chosen, and its affect on each term in the formal series understood. This is the approach taken by Edwards and Vilgis \cite{Edwards1986TheTransition}. The other is to include dynamic density fluctuations, as was done by Sussman and Schweizer \cite{Sussman2011MicroscopicMacromolecules}. While we expect our results to hold below the critical concentration, to truly understand the behaviour above it an approach similar to either of these is required.
	
	When interactions between the particle and more constraints are considered, our results may be ``washed out''. This was seen in a similar model problem; the diffusion of particles in a medium of static traps. Using a self consistent method similar to that laid out here, Bixon and Zwanzig \cite{Bixon1981DiffusionTraps} found interesting long time power-law tails in the concentration. When interactions with more traps were considered by Kirkpatrick \cite{Kirkpatrick1982TimeTraps}, these tails were found to cancel exactly. Ultimately, Grassberger and Procaccia \cite{Grassberger1982TheTraps} found that the tails were actually stretched exponentials. This story demonstrates the importance of truly understanding the interactions between the particle and many constraints, because they may have profound effects on our results. Nevertheless, our work represents a first pass at the problem and has already found very intriguing results.  

\begin{acknowledgments}
I would like to thank Prof. Randall D. Kamien for his encouragement and careful reading of the manuscript. This work was supported by the Simons Investigator Grant \# 291825 from the Simons Foundation.
\end{acknowledgments}
	\appendix
	\section{One Constraint Green Function}
	\label{app:CrossTPGF}
	In this appendix we compute the one constraint self-consistent adjoint Green function, $\widetilde{G}^{\dagger}_1$, for a bent wire frame, satisfying (\ref{CrossTPSCSEqn}). Here we write $\widetilde{G}^{\dagger}_{1} = G$. This will be split into two parts; one where the constraint is in the narrow wedge between the two legs and one when it is in the wider. The solution in the first region we call $G_{<}$ and in the second $G_{>}$. We will only discuss $G_{<}$ in detail and state the result for $G_{>}$ since its derivation is the same. 
    
    For $G_{<}$ let us work in terms of the relative angle $\xi = \varphi - \phi - \chi /2$ so that the boundary conditions $G_{<}$ satisfies (\ref{Cross2PGFBCs1}) are 
	\begin{equation}
		\frac{\partial G_{<}}{\partial \xi}\Bigg\lvert_{\xi = -\chi/2} = 	\frac{\partial G_{<}}{\partial \xi}\Bigg\lvert_{\xi = \chi/2 } =0 .
	\end{equation}
	An expansion for $G_{<}$ satisfying these conditions is
	\begin{equation}
		\label{CrossAppGSeries}
		G_{<}(r,r'; \xi , \xi' | s) = \sum_{n} g_{\alpha_{n}}(r,r'|s) \cos(\alpha_n \xi) \cos(\alpha_n \xi') + g_{\beta_{n}}(r,r'|s) \sin(\beta_n \xi)\sin(\beta_n \xi'),
	\end{equation}
	where $\alpha_n = 2 n \pi / \chi$ and $\beta_{n} = (2n+1)\pi / \chi$ for integer $n$. The $g_{\alpha_n}$ satisfy 
	\begin{equation}
 \label{galphaeqn}
		s g_{\alpha_n} - \frac{\widetilde{\mathcal{D}}(s)}{r}\frac{\partial}{\partial r}\left(r \frac{\partial g_{\alpha_n}}{\partial r}\right)- \alpha_{n}^2 \left(\frac{\widetilde{\mathcal{D}}(s)}{r^2}+\widetilde{D}_r(s)\right)g_{\alpha_n} = \frac{1}{r} \delta(r -r'),
	\end{equation}
    and the $g_{\beta_n}$ satisfy the same equation but replacing $\alpha_n \to \beta_n$. To solve (\ref{galphaeqn}) observe that $g_{\alpha}$ admits an eigenfunction expansion of standard form 
	\begin{equation}
		g_{\alpha_n}(r,r'|s) = \sum_{m} \frac{1}{\lambda_{\alpha_n}(m)} \Lambda_{\alpha_n}^m(r|s) \Lambda_{\alpha_n}^m(r'|s).
	\end{equation}
	Here the eigenfunctions, $\Lambda_{\alpha_n}^m(r|s)$, satisfy 
	\begin{equation}
		\label{CrossAppEigEqn}
		\frac{\widetilde{\mathcal{D}}(s)}{r}\frac{\partial}{\partial r}\left(r \frac{\partial \Lambda_{\alpha_n}^m(r|s)}{\partial r}\right) + \left(\lambda_n(m) - s - \alpha_n^2\widetilde{D}_r(s) - \frac{\alpha_n^2 \widetilde{\mathcal{D}}(s)}{r^2}\right) \Lambda_{\alpha_n}^m(r|s) = 0,
	\end{equation}
	 with $\lambda_{n}(m)$ being eigenvalues, indexed by $m$. The eigenfunctions $\Lambda_{\alpha_n}^m$ satisfy the orthogonality relations 
	\begin{equation}
		\int_{0}^{L/2} dr \Lambda_{\alpha_n}^m(r|s) \Lambda_{\alpha_p}^m(r|s) = \delta_{m p}.
	\end{equation}
 	Equation (\ref{CrossAppEigEqn}) is Bessel's equation, so the eigenfunctions are Bessel functions. We deduce the eigenvalues by insisting that the eigenfunctions vanish for $r\geq L/2$. This yields
 	\begin{equation}
 		\label{CrossAppGn}
 	      g_{\alpha_n}(r,r'|s) = \frac{4}{L^2} \sum_{m=1}^{\infty} \frac{1}{A^2_{\alpha_n}(m)}\frac{ J_{\alpha_n}\left(\frac{2}{L}j_{\alpha_n}(m) r\right) J_{\alpha_n}\left(\frac{2}{L} j_{\alpha_n}(m) r'\right)}{s + \alpha_n^2 \widetilde{D}_r(s) + \frac{4 \widetilde{\mathcal{D}}(s)}{L^2} j_{\alpha_n}^2(m)},
 	\end{equation}
 	where $J_{\alpha_n}(x)$ is the Bessel function of the first kind with order $\alpha_n$, $j_{\alpha_n}(m)$ is its $m^{\text{th}}$ zero and the constants $A_{\alpha_n}(m)$ are
 	\begin{equation}
 		A_{\alpha_n}(m) = \left(\int_{0}^{1} dx \ x J_{\alpha_n}^2\left(j_{\alpha_n}(m) x\right)\right)^{1/2}.
 	\end{equation}
  The solution for $g_{\beta_n}$ is the same as (\ref{CrossAppGn}) but with $\alpha_n \to \beta_n$. 

  To obtain the solution $G_{>}$, we follow the same steps starting from the expansion 
  	\begin{equation}
		\label{CrossAppGSeries2}
		G_{>}(r,r'; \xi , \xi' | s) = \sum_{n} f_{\eta_{n}}(r,r'|s) \cos(\eta_n \xi) \cos(\eta_n \xi') + f_{\zeta_{n}}(r,r'|s) \sin(\zeta_n \xi)\sin(\zeta_n \xi'),
	\end{equation}
 with $\eta_n = 2n \pi / (2\pi - \chi)$ and $\zeta_n = (2n + 1)\pi/ (2\pi - \chi)$. The solutions for $f_{\eta_n}$ and $f_{\zeta_n}$ are essentially the same as (\ref{CrossAppGn}).
	\section{Diffusion Kernel}
	Here we compute the one constraint self-consistent diffusion kernel for the wire frame. First $\widetilde{\Delta}$ is found using $\widetilde{G}_1^{\dagger}$ and (\ref{SCDrFDT}). Notice from (\ref{Tdefns}) that it will only be the contributions to $G_{<}$ and $G_{>}$ which are \textit{odd} in $\xi$ which contribute. The result is
	\begin{equation}
 \begin{split}
     		\widetilde{\Delta}(s) &= \frac{4}{L^2}\int_{0}^{L/2} dr \int_{0}^{L/2} dr' \ r \ r' \ \sum_{n\in \text{Even}}\sum_{m=1}^{\infty} \frac{2}{A^2_{\beta_n}(m)} \frac{J_{\beta_n}\left(\frac{2}{L} j_{\beta_n}(m) r\right) J_{\beta_n}\left(\frac{2}{L} j_{\beta_n}(m) r'\right)}{s + \beta_n^2 \widetilde{D}_r(s) + \frac{4 \widetilde{\mathcal{D}}(s)}{L^2} j_{\beta_n}^2(m)} \\
       &+\frac{4}{L^2}\int_{0}^{L/2} dr \int_{0}^{L/2} dr' \ r \ r' \ \sum_{n\in \text{Even}}\sum_{m=1}^{\infty} \frac{2}{A^2_{\zeta_n}(m)} \frac{J_{\zeta_n}\left(\frac{2}{L} j_{\zeta_n}(m) r\right) J_{\zeta_n}\left(\frac{2}{L} j_{\zeta_n}(m) r'\right)}{s + \zeta_n^2 \widetilde{D}_r(s) + \frac{4 \widetilde{\mathcal{D}}(s)}{L^2} j_{\zeta_n}^2(m)}.
 \end{split}
	\end{equation}
	Defining $x = 2 r / L$ and 
	\begin{equation}
		B_{\beta_n}(m) = \int_{0}^{1} dx \ x J_{\beta_n}\left(j_{\beta_n}(m) x\right),
	\end{equation}
	and the same for $\beta_n \to \zeta_{n}$, we can write
	\begin{equation}
 \begin{split}
     		\widetilde{\Delta}(s) &= \frac{2 L^4}{2^4 \widetilde{\mathcal{D}}(s)} \sum_{n\in \text{even}}\sum_{m=1}^{\infty} \left(\frac{B_{\beta_n}(m)}{A_{\beta_n}(m)}\right)^2 \left(j_{\beta_n}^2(m) + \frac{\beta_n^2 L^2 \widetilde{D}_r(s)}{ \widetilde{\mathcal{D}}(s)}+\frac{L^2 s}{4 \widetilde{\mathcal{D}}(s)}\right)^{-1}\\
       &+\frac{2 L^4}{2^4 \widetilde{\mathcal{D}}(s)} \sum_{n\in \text{even}}\sum_{m=1}^{\infty} \left(\frac{B_{\zeta_n}(m)}{A_{\zeta_n}(m)}\right)^2 \left(j_{\zeta_n}^2(m) + \frac{\zeta_n^2 L^2 \widetilde{D}_r(s)}{ \widetilde{\mathcal{D}}(s)}+\frac{L^2 s}{4 \widetilde{\mathcal{D}}(s)}\right)^{-1}
 \end{split}
	\end{equation}
	which, when written compactly, is equation (\ref{Delta1crosses}) of the main text. This defines the constants $C_n(m)$, which decrease quickly in magnitude as both $n$ and $m$ increase, as well as $a_n$ in terms of $\beta_n$ and $\zeta_n$. 
	
	To get the final expression for the diffusion kernel we approximate it by just the first term in the sum, i.e. the term with the larges ratio of $B$ to $A$. This yields the equation 
	\begin{equation}
		\frac{1}{\widetilde{D}_r(s)} = \frac{1}{D_r} + \frac{\rho L^4}{\widetilde{\mathcal{D}}(s)} \frac{a}{b+ \frac{L^2 s}{4 \widetilde{\mathcal{D}}(s)}}. 
	\end{equation}
	where $2 a = \left(B_{\zeta_0}(1)/ A_{\zeta_0}(1)\right)^2 \approx 0.4$ which is the largest coefficient when evaluated numerically and $b \approx 3.2$ is the square of the first zero of the Bessel function appropriate Bessel function. Then, we suppose that $4 \widetilde{\mathcal{D}} \sim L^2 \widetilde{D}_r(s)$ to obtain
		\begin{equation}
			\label{CrossDrEqnApp}
		\frac{1}{\widetilde{D}_r(s)} = \frac{1}{D_r} + \frac{\rho L^2}{\widetilde{D}_r(s)} \frac{a}{b+ s/\widetilde{D}_r(s)},
	\end{equation}	
	as in (\ref{Delta1Approx}) of the main text. We find the numerical value of $c^* = b / a \approx 16$.
	
	If we define $d(s) = \widetilde{D}_r(s)/D_r$ and $c = \rho L^2$, then (\ref{CrossDrEqnApp}) can be cast as a quadratic equation 
	\begin{equation}
		d^2(s) - \left((1 - a/b c) - \frac{s}{b D_r}\right) d(s) - \frac{s}{b D_r} = 0. 
	\end{equation}
	This is solved easily, and we determine which root we should take by insisting $d(s) \geq 0$ when $s \to 0$. This yields equation (\ref{fullDr}) of the main text. 
	\section{Mean Squared Displacement}
	\label{app:CrossMSD}
	 We cannot use (\ref{fullDr}) to obtain a simple closed form for $\mathcal{M}(t)$. However, it may be written in terms of a single integral. Let us first consider the inverse Laplace transform of $d(s)$. We write it in terms of the variable $z = \tau_0 s + 1 + c/c^*$, so that
	\begin{equation}
		2 d(s) = 2 - z + \sqrt{z^2 - 4 c/c^*} = 2 + \psi(z). 
	\end{equation}
	Taking into account the shifted variable and that everything is only a function of $\tau_0 s$, the inverse Laplace transform may be found in standard tables \cite{Abramowitz1964HandbookTables}
	\begin{equation}
		2 d(t) = 2 \delta(t) + \frac{1}{\tau_0} e^{-(1 + c/c^*) t/\tau_0} \psi(t/\tau_0),
	\end{equation}
	with
	\begin{equation}
	\psi(t) = - 2 \sqrt{\frac{c}{c^*}} \frac{1}{t} I_1\left(2  \sqrt{\frac{c}{c^*}} t\right),
	\end{equation}
	$I_1(x)$ is the first order modified Bessel function of the first kind.
	
	To find $\mathcal{M}(t)$ we note that,
	\begin{equation}
		\mathcal{M}(t) = 2 D_r \int_{0}^{t} dt' \int_0^{t'} dt'' \ d(t'') = 2D_r \int_{0}^{t} dt' \ (t - t') d(t').
 	\end{equation}
 	where the second equality is a standard identity \cite{Abramowitz1964HandbookTables}. The final result is
	\begin{equation}
	\mathcal{M}(t) = 2D_r t - 2 \sqrt{\frac{c}{c^*}} \int_{0}^{D_r t} dx \ \left(\frac{D_r t}{x} -1\right) e^{-(1+c/c*) x} I_1\left(2 \sqrt{\frac{c}{c^*}}x\right).
	\end{equation}
    A more formal derivation of (\ref{MSDcrosses}) can be found using multiple integrations by parts to obtain an asymptotic series \cite{Copson1965}. The results are the same as in the main text.
	\section{Elastic Stress}
    \label{app:crossstress}
		Here we provide the details for the derivation of the elastic stress (\ref{StressCross}). The first step is to identify $\delta P$ from the Smoluchowski equation, 
	\begin{equation}
		\delta P = - \dot{\gamma} \sum_{i=1}^{N} \left(r_i \sin \varphi_i \cos \varphi_i \frac{\partial P}{\partial r_i} + \sin^2 \varphi_i \frac{\partial P}{\partial \varphi_i}\right) + \frac{\dot{\gamma}}{2} \frac{\partial}{\partial \phi}\bigg(\left(\sin^2\phi + \sin^2(\phi+\chi)\right) P \bigg).
	\end{equation}
	This is substituted into (\ref{deltaF}) and integrated by parts. When we do this, we must remember that the area element in polar co-ordinates depends on $r_i$. The result, in full detail, is 
	\begin{equation}
		\begin{split}
			\frac{\delta F}{k_B T} = &- \frac{1}{2}\dot{\gamma} \int d\phi \int \prod_{i=1}^{N} d\varphi_i dr_i \ P(\phi,\{\textbf{r}_i\};t) \left(\sin^2\phi + \sin^2(\phi+\chi)\right) \frac{\partial}{\partial \phi}  \left(r_i\log P/P_{\text{eq}} \right)\\
			&+\dot{\gamma}\int d\phi \int \prod_{i=1}^{N} d\varphi_i dr_i \ P(\phi,\{\textbf{r}_i\};t) \sum_{i=1}^{N} \frac{\partial}{\partial r_i} \left(r_i^2 \sin \varphi_i \cos \varphi_i \log P/P_{\text{eq}}\right) \\
			&- \dot{\gamma}\int d\phi \int \prod_{i=1}^{N} d\varphi_i dr_i \ P(\phi,\{\textbf{r}_i\};t) \sum_{i=1}^{N} \frac{\partial}{\partial \varphi_i} \left(r_i \sin^2 \varphi_i \log P/P_{\text{eq}}\right).
		\end{split}
	\end{equation}
	If we expand out the derivatives, the terms proportional to $\log P/P_{\text{eq}}$ from the second and third lines will cancel. Recalling the derivatives of the equilibrium distribution, the remaining terms at this point are
	\begin{equation}
 \begin{split}
  	\frac{\delta F}{k_B T} =& - \frac{1}{2}\dot{\gamma} \int \left(\sin^2\phi + \sin^2(\phi+\chi)\right) \frac{\partial P}{\partial \phi} + \frac{1}{2}\dot{\gamma} \int P \sum_i \left(\sin^2 \phi + \sin^2(\phi +\chi) - 2 \sin^2 \varphi_i\right) \mathcal{T}_i \Theta_i \\
   & -\dot{\gamma} \sum_i \int   \left(\sin^2 \varphi_i \frac{\partial P}{\partial \varphi_i} - r_i \sin \varphi_i \cos \varphi_i \frac{\partial P}{\partial r_i} \right).
 \end{split}
	\end{equation}
	here we are fee to use our shorthand because we are integrating over the correct area element. For the last term, one must integrate by parts making sure that the derivative with respect to $r_i$ acts on the area element. Then it is straightforward to show that this vanishes. The second term is already that appearing in (\ref{StressCross}) and the first term requires one last integration by parts before we finally obtain the elastic stress as in (\ref{StressCross}) of the main text.
 \section{Simplifying the relaxation modulus}
 \label{app:relax}
    Here we give some details of how the relaxation modulus can be simplified to give (\ref{fullmod}). First we consider the internal modulus. Using the identity $2\sin x \sin y = \cos(x - y)- \cos(x+y)$, we can write this as
    \begin{equation}
        \Gamma_{\text{Int}}(t) = \frac{1 }{2} k_B T \cos^2 \chi \left(\left\langle \cos\left[2 (\phi(t)-\phi(0)) \right] \right\rangle_{\text{eq}} - \left\langle \cos\left[2 (\phi(t) + \phi(0) + \chi) \right] \right\rangle_{\text{eq}}\right).
    \end{equation}
    The second equilibrium average in the above can be written
    \begin{equation}
    \label{cancel}
        \left\langle \cos\left[2 (\phi(t) + \phi(0) + \chi) \right] \right\rangle_{\text{eq}} = \left\langle \cos^2 (\phi(t) + \phi(0) + \chi) \right\rangle_{\text{eq}} - \left\langle \sin^2 (\phi(t) + \phi(0) + \chi) \right\rangle_{\text{eq}}. 
    \end{equation}
    Notice that the second term here is equal to the first, but with both the initial and final orientations, $\phi(0)$ and $\phi(t)$, rotated by $\pi/2$. These terms must therefore \textit{cancel}, since these averages are taken at an assumed isotropic equilibrium. The internal modulus is then as given in (\ref{IntMod}) of the main text.
    
    The external modulus requires a little more work. First let us inspect $S_{\text{Ext}}$. By using the delta functions in $\mathcal{T}_i$ we can re-write this as
    \begin{equation}
    \label{SExtMan}
        S_{\text{Ext}} = \sum_{i=1}^{N} \left(\sin^2(\phi+\chi) - \sin^2(\phi)\right) \left(\delta(\varphi_i - \phi) + \delta(\varphi_i - \phi - \chi)\right) \Theta(L/2-r_i). 
    \end{equation}
    We define $\mathcal{T}^{\dagger}_i = \delta(\varphi_i - \phi) + \delta(\varphi_i - \phi - \chi)$. When we use (\ref{SExtMan}) in (\ref{GammaExt}) with the help of double angle formulae and the same identity as before we find several terms. Some of these are functions of the difference $\phi(t)-\phi(0)$, and some depend on $\phi(t)+\phi(0)$. The same reasons which caused the cancellation in (\ref{cancel}) lead to all those in the latter category to vanish. Hence we find the external modulus is as given in (\ref{ExtMod}).
    
    Similar manipulations can be used to obtain the cross term as given in (\ref{MixedMod}).
 \section{Recursion relation for \texorpdfstring{$\mathcal{M}_n$}{TEXT}}
\label{app:rec}
In this appendix we derive the recursion relation for the moments $\mathcal{M}_n(t) = \langle (\phi(t)-\phi(0))^n\rangle_{\text{eq}}$. It is most convenient to work in Laplace transform so that this may be written
\begin{equation}
\label{Mnapp}
    \mathcal{M}_n(s) = \int d\phi \prod_{i=1}^{N} d\textbf{r}_i \int d\phi' \prod_{i=1}^{N} d\textbf{r}'_i \ (\phi-\phi')^n \widetilde{G}(\phi, \{\textbf{r}_i\};\phi',\{\textbf{r}'_i\}|s) \frac{1}{s} P_{\text{eq}}(\phi',\{\textbf{r}'_i\})  
\end{equation}
where $\widetilde{G}$ is the Green function satisfying
\begin{equation}
\label{Gapp}
    s \widetilde{G}(s) = \widetilde{D}_r(s) \frac{\partial^2 \widetilde{G}(s)}{\partial \phi^2} + \sum_{i=1}^{N} \nabla^2_i \widetilde{G}(s) + \delta(\phi - \phi') \prod_{i=1}^{N} \delta(\textbf{r}_i-\textbf{r}'_i). 
\end{equation}
Multiplying (\ref{Mnapp}) by $s$ (i.e. taking its time derivative) and making use of (\ref{Gapp}) yields, in shorthand
\begin{equation}
    s \mathcal{M}_n(s) = \int \int  \ (\phi-\phi')^n \left[\widetilde{D}_r(s) \frac{\partial^2 \widetilde{G}(s)}{\partial \phi^2} + \sum_{i=1}^{N} \nabla^2_i \widetilde{G}(s)\right] \frac{1}{s} P_{\text{eq}}.
\end{equation}
Two integrations by parts yield 
\begin{equation}
    s \mathcal{M}_n(s) = n (n-1) \widetilde{D}_r(s)\int \int  \ (\phi-\phi')^{n-2} \widetilde{G}(s) \frac{1}{s} P_{\text{eq}}
\end{equation}
which, assuming $n >2$ implies (\ref{Mnrec}). The same manipulations can be used to prove the initial conditions for the recursion relation (\ref{Mnrecint}).
\section{Time dependence of the stress}
\label{app:time}
Here we show how the long and short time limits of $R(t)$ may be extracted from (\ref{Rs}) and (\ref{fullDr}). First, we discuss the short time limit, $t \ll \tau_0$. Taking $s\tau_0 \gg 1 $ in (\ref{fullDr}) yields
\begin{equation}
    \widetilde{D}_r(s) \sim D_r + \mathcal{O}((\tau_0 s)^{-1}). 
\end{equation}
Hence, 
\begin{equation}
    R(s) \sim \frac{1}{s}\sum_{n=0}^{\infty} (-1)^n \left(\frac{4 D_r}{s}\right)^n.
\end{equation}
Then we can invert the Laplace transform term by term to give
\begin{equation}
    R(t) \sim \frac{1}{s}\sum_{n=0}^{\infty} \frac{(-1)^n}{n!} \left( D_r t\right)^n = e^{-4 D_r t},
\end{equation}
as in (\ref{Rtshort}) of the main text.

For the long time limit we need to consider the limit $s \tau_0 \ll 1$. When $c< c^*$ we find
\begin{equation}
    \widetilde{D}_r(s) \sim D_r (1- c/c^*) + \mathcal{O}(\tau_0 s). 
\end{equation}
The same steps as above then lead to
\begin{equation}
    R(t) \sim e^{-4 D_r(1- c/c^*) t}.
\end{equation}
Precisely when $c = c^*$ equation (\ref{fullDr}) gives
\begin{equation}
   \widetilde{D}_r(s) \sim \sqrt{D_r s} + \mathcal{O}(\tau_0 s). 
\end{equation}
So we have
\begin{equation}
    R(s) \sim \frac{1}{s}\sum_{n=0}^{\infty} (-1)^n \left(4 \sqrt{\frac{D_r}{s}}\right)^n,
\end{equation}
to which we again apply the inverse Laplace transform term by term 
\begin{equation}
    R(t) \sim \sum_{n=0}^{\infty} (-1)^n \frac{4 \sqrt{D_r t}}{\Gamma(n/2 +1)}.
\end{equation}
This infinite series can now be written in terms of an error function \cite{Abramowitz1964HandbookTables}
\begin{equation}
    R(t) \sim e^{16 D_r t} \left(1 - \text{erf}(4 \sqrt{D_r t})\right).
\end{equation}
Now we recall that this result should only be valid for $D_r t \gg 1$ for which the above becomes 
\begin{equation}
    R(t) \sim \frac{1}{4\sqrt{\pi D_r t}}.
\end{equation}
Finally we consider $c>c^*$. In this case the short time limit of (\ref{fullDr}) is 
\begin{equation}
    \widetilde{D}_r(s) \sim \frac{s}{|1-c/c^*|} + \mathcal{O}((s \tau_0)^2),
\end{equation}
whence 
\begin{equation}
    R(s) \sim \frac{1}{s}\sum_{n=0}^{\infty} (-1)^n \left(\frac{4}{|1-c/c^*|}\right)^n,
\end{equation}
then after inverting the Laplace transform
\begin{equation}
    R(t) \sim \sum_{n=0}^{\infty} (-1)^n \left(\frac{4}{|1-c/c^*|}\right)^n = \frac{c/c^* -1}{c/c^* +3},
\end{equation}
where the final equality holds for sufficiently large $c/c^*$. Collating these results yields (\ref{rt}).
\bibliography{references}

\begin{thebibliography}{70}%
\makeatletter
\providecommand \@ifxundefined [1]{%
 \@ifx{#1\undefined}
}%
\providecommand \@ifnum [1]{%
 \ifnum #1\expandafter \@firstoftwo
 \else \expandafter \@secondoftwo
 \fi
}%
\providecommand \@ifx [1]{%
 \ifx #1\expandafter \@firstoftwo
 \else \expandafter \@secondoftwo
 \fi
}%
\providecommand \natexlab [1]{#1}%
\providecommand \enquote  [1]{``#1''}%
\providecommand \bibnamefont  [1]{#1}%
\providecommand \bibfnamefont [1]{#1}%
\providecommand \citenamefont [1]{#1}%
\providecommand \href@noop [0]{\@secondoftwo}%
\providecommand \href [0]{\begingroup \@sanitize@url \@href}%
\providecommand \@href[1]{\@@startlink{#1}\@@href}%
\providecommand \@@href[1]{\endgroup#1\@@endlink}%
\providecommand \@sanitize@url [0]{\catcode `\\12\catcode `\$12\catcode
  `\&12\catcode `\#12\catcode `\^12\catcode `\_12\catcode `\%12\relax}%
\providecommand \@@startlink[1]{}%
\providecommand \@@endlink[0]{}%
\providecommand \url  [0]{\begingroup\@sanitize@url \@url }%
\providecommand \@url [1]{\endgroup\@href {#1}{\urlprefix }}%
\providecommand \urlprefix  [0]{URL }%
\providecommand \Eprint [0]{\href }%
\providecommand \doibase [0]{https://doi.org/}%
\providecommand \selectlanguage [0]{\@gobble}%
\providecommand \bibinfo  [0]{\@secondoftwo}%
\providecommand \bibfield  [0]{\@secondoftwo}%
\providecommand \translation [1]{[#1]}%
\providecommand \BibitemOpen [0]{}%
\providecommand \bibitemStop [0]{}%
\providecommand \bibitemNoStop [0]{.\EOS\space}%
\providecommand \EOS [0]{\spacefactor3000\relax}%
\providecommand \BibitemShut  [1]{\csname bibitem#1\endcsname}%
\let\auto@bib@innerbib\@empty
\bibitem [{\citenamefont {Einstein}(1905)}]{Einstein1905UberTeilchen}%
  \BibitemOpen
  \bibfield  {author} {\bibinfo {author} {\bibfnamefont {A.}~\bibnamefont
  {Einstein}},\ }\bibfield  {title} {\bibinfo {title} {{{\"{U}}ber die von der
  molekularkinetischen Theorie der W{\"{a}}rme geforderte Bewegung von in
  ruhenden Fl{\"{u}}ssigkeiten suspendierten Teilchen}},\ }\href@noop {}
  {\bibfield  {journal} {\bibinfo  {journal} {Annalen der Physik}\ }\textbf
  {\bibinfo {volume} {322}},\ \bibinfo {pages} {549} (\bibinfo {year}
  {1905})}\BibitemShut {NoStop}%
\bibitem [{\citenamefont {Dhont}(1996)}]{Dhont1996AnColloids}%
  \BibitemOpen
  \bibfield  {author} {\bibinfo {author} {\bibfnamefont {J.~K.~G.}\
  \bibnamefont {Dhont}},\ }\href@noop {} {\emph {\bibinfo {title} {{An
  Introduction to Dynamics of Colloids}}}}\ (\bibinfo  {publisher} {Elsevier},\
  \bibinfo {year} {1996})\BibitemShut {NoStop}%
\bibitem [{\citenamefont {Leegwater}\ and\ \citenamefont
  {Szamel}(1992{\natexlab{a}})}]{Leegwater1992Long-Fluids}%
  \BibitemOpen
  \bibfield  {author} {\bibinfo {author} {\bibfnamefont {J.~A.}\ \bibnamefont
  {Leegwater}}\ and\ \bibinfo {author} {\bibfnamefont {G.}~\bibnamefont
  {Szamel}},\ }\bibfield  {title} {\bibinfo {title} {{Long- and ultimate-time
  tails in two-dimensional fluids}},\ }\href@noop {} {\bibfield  {journal}
  {\bibinfo  {journal} {Physical Review A}\ }\textbf {\bibinfo {volume} {45}},\
  \bibinfo {pages} {1270} (\bibinfo {year} {1992}{\natexlab{a}})}\BibitemShut
  {NoStop}%
\bibitem [{\citenamefont {Leegwater}\ and\ \citenamefont
  {Szamel}(1992{\natexlab{b}})}]{Leegwater1992DynamicalSuspensions}%
  \BibitemOpen
  \bibfield  {author} {\bibinfo {author} {\bibfnamefont {J.~A.}\ \bibnamefont
  {Leegwater}}\ and\ \bibinfo {author} {\bibfnamefont {G.}~\bibnamefont
  {Szamel}},\ }\bibfield  {title} {\bibinfo {title} {{Dynamical properties of
  hard-sphere suspensions}},\ }\href {https://doi.org/10.1103/PhysRevA.46.4999}
  {\bibfield  {journal} {\bibinfo  {journal} {Physical Review A}\ }\textbf
  {\bibinfo {volume} {46}},\ \bibinfo {pages} {4999} (\bibinfo {year}
  {1992}{\natexlab{b}})}\BibitemShut {NoStop}%
\bibitem [{\citenamefont {Szamel}\ and\ \citenamefont
  {Leegwater}(1992)}]{Szamel1992Long-timeSuspensions}%
  \BibitemOpen
  \bibfield  {author} {\bibinfo {author} {\bibfnamefont {G.}~\bibnamefont
  {Szamel}}\ and\ \bibinfo {author} {\bibfnamefont {J.~A.}\ \bibnamefont
  {Leegwater}},\ }\bibfield  {title} {\bibinfo {title} {{Long-time
  self-diffusion coefficients of suspensions}},\ }\href
  {https://doi.org/10.1103/PhysRevA.46.5012} {\bibfield  {journal} {\bibinfo
  {journal} {Physical Review A}\ }\textbf {\bibinfo {volume} {46}},\ \bibinfo
  {pages} {5012} (\bibinfo {year} {1992})}\BibitemShut {NoStop}%
\bibitem [{\citenamefont
  {Einstein}(1906)}]{Einstein1906EineMolekul-dimensionen}%
  \BibitemOpen
  \bibfield  {author} {\bibinfo {author} {\bibfnamefont {A.}~\bibnamefont
  {Einstein}},\ }\bibfield  {title} {\bibinfo {title} {{Eine neue Bestimmung
  der Molek{\"{u}}l-dimensionen}},\ }\href@noop {} {\bibfield  {journal}
  {\bibinfo  {journal} {Annalen der Physik}\ }\textbf {\bibinfo {volume}
  {iv}},\ \bibinfo {pages} {289} (\bibinfo {year} {1906})}\BibitemShut
  {NoStop}%
\bibitem [{\citenamefont
  {Einstein}(1911)}]{Einstein1911BerichtigungMolekul-dimensionen}%
  \BibitemOpen
  \bibfield  {author} {\bibinfo {author} {\bibfnamefont {A.}~\bibnamefont
  {Einstein}},\ }\bibfield  {title} {\bibinfo {title} {{Berichtigung zu meiner
  Arbeit: Eine neue Bestimmung der Molek{\"{u}}l-dimensionen}},\ }\href@noop {}
  {\bibfield  {journal} {\bibinfo  {journal} {Annalen der Physik}\ }\textbf
  {\bibinfo {volume} {iv}},\ \bibinfo {pages} {591} (\bibinfo {year}
  {1911})}\BibitemShut {NoStop}%
\bibitem [{\citenamefont {Batchelor}\ and\ \citenamefont
  {Green}(1972)}]{Batchelor1972TheC2}%
  \BibitemOpen
  \bibfield  {author} {\bibinfo {author} {\bibfnamefont {G.~K.}\ \bibnamefont
  {Batchelor}}\ and\ \bibinfo {author} {\bibfnamefont {J.~T.}\ \bibnamefont
  {Green}},\ }\bibfield  {title} {\bibinfo {title} {{The determination of the
  bulk stress in a suspension of spherical particles to order c{\^{}}2}},\
  }\href {http://sor.scitation.org/doi/10.1122/1.548848} {\bibfield  {journal}
  {\bibinfo  {journal} {Journal of Fluid Mechanics}\ }\textbf {\bibinfo
  {volume} {56}},\ \bibinfo {pages} {401} (\bibinfo {year} {1972})}\BibitemShut
  {NoStop}%
\bibitem [{\citenamefont {Trappe}\ \emph {et~al.}(2001)\citenamefont {Trappe},
  \citenamefont {Prasad}, \citenamefont {Cipelletti}, \citenamefont {Segre},\
  and\ \citenamefont {Weitz}}]{Trappe2001}%
  \BibitemOpen
  \bibfield  {author} {\bibinfo {author} {\bibfnamefont {V.}~\bibnamefont
  {Trappe}}, \bibinfo {author} {\bibfnamefont {V.}~\bibnamefont {Prasad}},
  \bibinfo {author} {\bibfnamefont {L.}~\bibnamefont {Cipelletti}}, \bibinfo
  {author} {\bibfnamefont {P.~N.}\ \bibnamefont {Segre}},\ and\ \bibinfo
  {author} {\bibfnamefont {D.~A.}\ \bibnamefont {Weitz}},\ }\bibfield  {title}
  {\bibinfo {title} {{Jamming phase diagram for attractive particles}},\ }\href
  {https://doi.org/10.1038/35081021} {\bibfield  {journal} {\bibinfo  {journal}
  {Nature}\ }\textbf {\bibinfo {volume} {411}},\ \bibinfo {pages} {772}
  (\bibinfo {year} {2001})}\BibitemShut {NoStop}%
\bibitem [{\citenamefont {Happel}\ and\ \citenamefont
  {Brenner}(1973)}]{Happel1973LowMedia.}%
  \BibitemOpen
  \bibfield  {author} {\bibinfo {author} {\bibfnamefont {J.}~\bibnamefont
  {Happel}}\ and\ \bibinfo {author} {\bibfnamefont {H.}~\bibnamefont
  {Brenner}},\ }\href
  {https://books.google.co.uk/books/about/Low_Reynolds_Number_Hydrodynamics.html?id=LLe-QgAACAAJ&source=kp_book_description&redir_esc=y}
  {\emph {\bibinfo {title} {{Low Reynolds number hydrodynamics. With special
  applications to particulate media.}}}}\ (\bibinfo  {publisher} {Noordhoff
  International Pub},\ \bibinfo {year} {1973})\BibitemShut {NoStop}%
\bibitem [{\citenamefont {Kim}\ and\ \citenamefont
  {Karrila}(2005)}]{Kim2005MicrohydrodynamicsApplications}%
  \BibitemOpen
  \bibfield  {author} {\bibinfo {author} {\bibfnamefont {S.}~\bibnamefont
  {Kim}}\ and\ \bibinfo {author} {\bibfnamefont {S.~J.}\ \bibnamefont
  {Karrila}},\ }\href
  {https://books.google.co.uk/books?hl=en&lr=&id=aADEAgAAQBAJ&oi=fnd&pg=PP1&dq=kim+karrila+microhydrodynamics&ots=hH91igZ3Mr&sig=XVf21MGBQXz5H7d7iDv-LrkDgJg#v=onepage&q=kim
  karrila microhydrodynamics&f=false} {\emph {\bibinfo {title}
  {{Microhydrodynamics : principles and selected applications}}}}\ (\bibinfo
  {publisher} {Dover Publications},\ \bibinfo {year} {2005})\BibitemShut
  {NoStop}%
\bibitem [{\citenamefont {Makino}\ and\ \citenamefont
  {Doi}(2004{\natexlab{a}})}]{Makino2004BrownianFluid}%
  \BibitemOpen
  \bibfield  {author} {\bibinfo {author} {\bibfnamefont {M.}~\bibnamefont
  {Makino}}\ and\ \bibinfo {author} {\bibfnamefont {M.}~\bibnamefont {Doi}},\
  }\bibfield  {title} {\bibinfo {title} {{Brownian motion of a particle of
  general shape in Newtonian fluid}},\ }\href
  {https://doi.org/10.1143/JPSJ.73.2739} {\bibfield  {journal} {\bibinfo
  {journal} {Journal of the Physical Society of Japan}\ }\textbf {\bibinfo
  {volume} {73}},\ \bibinfo {pages} {2739} (\bibinfo {year}
  {2004}{\natexlab{a}})}\BibitemShut {NoStop}%
\bibitem [{\citenamefont {Makino}\ and\ \citenamefont
  {Doi}(2004{\natexlab{b}})}]{Makino2004ViscoelasticityShape}%
  \BibitemOpen
  \bibfield  {author} {\bibinfo {author} {\bibfnamefont {M.}~\bibnamefont
  {Makino}}\ and\ \bibinfo {author} {\bibfnamefont {M.}~\bibnamefont {Doi}},\
  }\bibfield  {title} {\bibinfo {title} {{Viscoelasticity of dilute solutions
  of particles of general shape}},\ }\href
  {https://doi.org/10.1143/JPSJ.73.3020} {\bibfield  {journal} {\bibinfo
  {journal} {Journal of the Physical Society of Japan}\ }\textbf {\bibinfo
  {volume} {73}},\ \bibinfo {pages} {3020} (\bibinfo {year}
  {2004}{\natexlab{b}})}\BibitemShut {NoStop}%
\bibitem [{\citenamefont {Dubin}\ \emph {et~al.}(1967)\citenamefont {Dubin},
  \citenamefont {Lunacek},\ and\ \citenamefont {Benedek}}]{Dubin1967}%
  \BibitemOpen
  \bibfield  {author} {\bibinfo {author} {\bibfnamefont {S.~B.}\ \bibnamefont
  {Dubin}}, \bibinfo {author} {\bibfnamefont {J.~H.}\ \bibnamefont {Lunacek}},\
  and\ \bibinfo {author} {\bibfnamefont {G.~B.}\ \bibnamefont {Benedek}},\
  }\bibfield  {title} {\bibinfo {title} {{Observation of the spectrum of light
  scattered by solutions of biological macromolecules.}},\ }\href
  {https://doi.org/10.1073/pnas.57.5.1164} {\bibfield  {journal} {\bibinfo
  {journal} {Proceedings of the National Academy of Sciences of the United
  States of America}\ }\textbf {\bibinfo {volume} {57}},\ \bibinfo {pages}
  {1164} (\bibinfo {year} {1967})}\BibitemShut {NoStop}%
\bibitem [{\citenamefont {Nemoto}\ \emph {et~al.}(1975)\citenamefont {Nemoto},
  \citenamefont {Schrag}, \citenamefont {Ferry},\ and\ \citenamefont
  {Fulton}}]{Nemoto1975}%
  \BibitemOpen
  \bibfield  {author} {\bibinfo {author} {\bibfnamefont {N.}~\bibnamefont
  {Nemoto}}, \bibinfo {author} {\bibfnamefont {J.~L.}\ \bibnamefont {Schrag}},
  \bibinfo {author} {\bibfnamefont {J.~D.}\ \bibnamefont {Ferry}},\ and\
  \bibinfo {author} {\bibfnamefont {R.~W.}\ \bibnamefont {Fulton}},\ }\bibfield
   {title} {\bibinfo {title} {{Infinite‐dilution viscoelastic properties of
  tobacco mosaic virus}},\ }\href {https://doi.org/10.1002/bip.1975.360140213}
  {\bibfield  {journal} {\bibinfo  {journal} {Biopolymers}\ }\textbf {\bibinfo
  {volume} {14}},\ \bibinfo {pages} {409} (\bibinfo {year} {1975})}\BibitemShut
  {NoStop}%
\bibitem [{\citenamefont {Nicolai}\ and\ \citenamefont
  {Cocard}(2000)}]{Nicolai2000}%
  \BibitemOpen
  \bibfield  {author} {\bibinfo {author} {\bibfnamefont {T.}~\bibnamefont
  {Nicolai}}\ and\ \bibinfo {author} {\bibfnamefont {S.}~\bibnamefont
  {Cocard}},\ }\bibfield  {title} {\bibinfo {title} {{Light scattering study of
  the dispersion of laponite}},\ }\href {https://doi.org/10.1021/la9915623}
  {\bibfield  {journal} {\bibinfo  {journal} {Langmuir}\ }\textbf {\bibinfo
  {volume} {16}},\ \bibinfo {pages} {8189} (\bibinfo {year}
  {2000})}\BibitemShut {NoStop}%
\bibitem [{\citenamefont {Seeman}(1982)}]{Seeman1982NucleicLatticesb}%
  \BibitemOpen
  \bibfield  {author} {\bibinfo {author} {\bibfnamefont {N.~C.}\ \bibnamefont
  {Seeman}},\ }\bibfield  {title} {\bibinfo {title} {{Nucleic acid junctions
  and lattices}},\ }\href {https://doi.org/10.1016/0022-5193(82)90002-9}
  {\bibfield  {journal} {\bibinfo  {journal} {Journal of Theoretical Biology}\
  }\textbf {\bibinfo {volume} {99}},\ \bibinfo {pages} {237} (\bibinfo {year}
  {1982})}\BibitemShut {NoStop}%
\bibitem [{\citenamefont {Rothemund}(2006)}]{Rothemund2006FoldingPatterns}%
  \BibitemOpen
  \bibfield  {author} {\bibinfo {author} {\bibfnamefont {P.~W.}\ \bibnamefont
  {Rothemund}},\ }\bibfield  {title} {\bibinfo {title} {{Folding DNA to create
  nanoscale shapes and patterns}},\ }\href
  {https://doi.org/10.1038/nature04586} {\bibfield  {journal} {\bibinfo
  {journal} {Nature}\ }\textbf {\bibinfo {volume} {440}},\ \bibinfo {pages}
  {297} (\bibinfo {year} {2006})}\BibitemShut {NoStop}%
\bibitem [{\citenamefont {Han}\ \emph {et~al.}(2011)\citenamefont {Han},
  \citenamefont {Pal}, \citenamefont {Nangreave}, \citenamefont {Deng},
  \citenamefont {Liu},\ and\ \citenamefont {Yan}}]{Han2011DNASpace}%
  \BibitemOpen
  \bibfield  {author} {\bibinfo {author} {\bibfnamefont {D.}~\bibnamefont
  {Han}}, \bibinfo {author} {\bibfnamefont {S.}~\bibnamefont {Pal}}, \bibinfo
  {author} {\bibfnamefont {J.}~\bibnamefont {Nangreave}}, \bibinfo {author}
  {\bibfnamefont {Z.}~\bibnamefont {Deng}}, \bibinfo {author} {\bibfnamefont
  {Y.}~\bibnamefont {Liu}},\ and\ \bibinfo {author} {\bibfnamefont
  {H.}~\bibnamefont {Yan}},\ }\bibfield  {title} {\bibinfo {title} {{DNA
  origami with complex curvatures in three-dimensional space}},\ }\href
  {https://doi.org/10.1126/science.1202998} {\bibfield  {journal} {\bibinfo
  {journal} {Science}\ }\textbf {\bibinfo {volume} {332}},\ \bibinfo {pages}
  {342} (\bibinfo {year} {2011})}\BibitemShut {NoStop}%
\bibitem [{\citenamefont {Castro}\ \emph {et~al.}(2011)\citenamefont {Castro},
  \citenamefont {Kilchherr}, \citenamefont {Kim}, \citenamefont {Shiao},
  \citenamefont {Wauer}, \citenamefont {Wortmann}, \citenamefont {Bathe},\ and\
  \citenamefont {Dietz}}]{Castro2011AOrigami}%
  \BibitemOpen
  \bibfield  {author} {\bibinfo {author} {\bibfnamefont {C.~E.}\ \bibnamefont
  {Castro}}, \bibinfo {author} {\bibfnamefont {F.}~\bibnamefont {Kilchherr}},
  \bibinfo {author} {\bibfnamefont {D.~N.}\ \bibnamefont {Kim}}, \bibinfo
  {author} {\bibfnamefont {E.~L.}\ \bibnamefont {Shiao}}, \bibinfo {author}
  {\bibfnamefont {T.}~\bibnamefont {Wauer}}, \bibinfo {author} {\bibfnamefont
  {P.}~\bibnamefont {Wortmann}}, \bibinfo {author} {\bibfnamefont
  {M.}~\bibnamefont {Bathe}},\ and\ \bibinfo {author} {\bibfnamefont
  {H.}~\bibnamefont {Dietz}},\ }\bibfield  {title} {\bibinfo {title} {{A primer
  to scaffolded DNA origami}},\ }\href {https://doi.org/10.1038/nmeth.1570}
  {\bibfield  {journal} {\bibinfo  {journal} {Nature Methods}\ }\textbf
  {\bibinfo {volume} {8}},\ \bibinfo {pages} {221} (\bibinfo {year}
  {2011})}\BibitemShut {NoStop}%
\bibitem [{\citenamefont
  {Edwards}(1967{\natexlab{a}})}]{Edwards1967StatisticalI}%
  \BibitemOpen
  \bibfield  {author} {\bibinfo {author} {\bibfnamefont {S.~F.}\ \bibnamefont
  {Edwards}},\ }\bibfield  {title} {\bibinfo {title} {{Statistical mechanics
  with topological constraints: I}},\ }\href
  {https://doi.org/10.1088/0370-1328/91/3/301} {\bibfield  {journal} {\bibinfo
  {journal} {Proceedings of the Physical Society}\ }\textbf {\bibinfo {volume}
  {91}},\ \bibinfo {pages} {513} (\bibinfo {year}
  {1967}{\natexlab{a}})}\BibitemShut {NoStop}%
\bibitem [{\citenamefont
  {Edwards}(1967{\natexlab{b}})}]{Edwards1967StatisticalII}%
  \BibitemOpen
  \bibfield  {author} {\bibinfo {author} {\bibfnamefont {S.~F.}\ \bibnamefont
  {Edwards}},\ }\bibfield  {title} {\bibinfo {title} {{Statistical mechanics
  with topological constraints: II}},\ }\href
  {https://doi.org/10.1088/0370-1328/91/3/301} {\bibfield  {journal} {\bibinfo
  {journal} {Proceedings of the Physical Society}\ }\textbf {\bibinfo {volume}
  {91}},\ \bibinfo {pages} {513} (\bibinfo {year}
  {1967}{\natexlab{b}})}\BibitemShut {NoStop}%
\bibitem [{\citenamefont {Doi}\ and\ \citenamefont
  {Edwards}(1986)}]{Doi1986TheDynamics}%
  \BibitemOpen
  \bibfield  {author} {\bibinfo {author} {\bibfnamefont {M.}~\bibnamefont
  {Doi}}\ and\ \bibinfo {author} {\bibfnamefont {S.~F.}\ \bibnamefont
  {Edwards}},\ }\href
  {https://books.google.co.uk/books/about/The_Theory_of_Polymer_Dynamics.html?id=dMzGyWs3GKcC}
  {\emph {\bibinfo {title} {{The theory of polymer dynamics}}}}\ (\bibinfo
  {publisher} {Oxford University Press},\ \bibinfo {year} {1986})\BibitemShut
  {NoStop}%
\bibitem [{\citenamefont {Onsager}(1949)}]{Onsager1949TheParticles}%
  \BibitemOpen
  \bibfield  {author} {\bibinfo {author} {\bibfnamefont {L.}~\bibnamefont
  {Onsager}},\ }\bibfield  {title} {\bibinfo {title} {{the Effects of Shape on
  the Interaction of Colloidal Particles}},\ }\href
  {https://doi.org/10.1111/j.1749-6632.1949.tb27296.x} {\bibfield  {journal}
  {\bibinfo  {journal} {Annals of the New York Academy of Sciences}\ }\textbf
  {\bibinfo {volume} {51}},\ \bibinfo {pages} {627} (\bibinfo {year}
  {1949})}\BibitemShut {NoStop}%
\bibitem [{\citenamefont {Doi}(1975)}]{Doi1975RotationalSolution}%
  \BibitemOpen
  \bibfield  {author} {\bibinfo {author} {\bibfnamefont {M.}~\bibnamefont
  {Doi}},\ }\bibfield  {title} {\bibinfo {title} {{Rotational relaxation time
  of rigid rod-like macromolecule in concentrated solution}},\ }\href
  {https://doi.org/10.1051/jphys:01975003607-8060700} {\bibfield  {journal}
  {\bibinfo  {journal} {Journal de Physique}\ }\textbf {\bibinfo {volume}
  {36}},\ \bibinfo {pages} {607} (\bibinfo {year} {1975})}\BibitemShut
  {NoStop}%
\bibitem [{\citenamefont {Doi}\ and\ \citenamefont
  {Edwards}(1978{\natexlab{a}})}]{Doi1978Dynamics1}%
  \BibitemOpen
  \bibfield  {author} {\bibinfo {author} {\bibfnamefont {M.}~\bibnamefont
  {Doi}}\ and\ \bibinfo {author} {\bibfnamefont {S.~F.}\ \bibnamefont
  {Edwards}},\ }\bibfield  {title} {\bibinfo {title} {{Dynamics of rod-like
  macromolecules in concentrated solution. Part 1}},\ }\href
  {https://doi.org/10.1039/f29787400560} {\bibfield  {journal} {\bibinfo
  {journal} {Journal of the Chemical Society, Faraday Transactions 2}\ }\textbf
  {\bibinfo {volume} {74}},\ \bibinfo {pages} {560} (\bibinfo {year}
  {1978}{\natexlab{a}})}\BibitemShut {NoStop}%
\bibitem [{\citenamefont {Doi}\ and\ \citenamefont
  {Edwards}(1978{\natexlab{b}})}]{Doi1978Dynamics2}%
  \BibitemOpen
  \bibfield  {author} {\bibinfo {author} {\bibfnamefont {M.}~\bibnamefont
  {Doi}}\ and\ \bibinfo {author} {\bibfnamefont {S.~F.}\ \bibnamefont
  {Edwards}},\ }\bibfield  {title} {\bibinfo {title} {{Dynamics of rod-like
  macromolecules in concentrated solution. Part 2}},\ }\href
  {https://doi.org/10.1039/f29787400918} {\bibfield  {journal} {\bibinfo
  {journal} {Journal of the Chemical Society, Faraday Transactions 2}\ }\textbf
  {\bibinfo {volume} {74}},\ \bibinfo {pages} {918} (\bibinfo {year}
  {1978}{\natexlab{b}})}\BibitemShut {NoStop}%
\bibitem [{\citenamefont {De~Gennes}(1971)}]{DeGennes1971ReptationObstacles}%
  \BibitemOpen
  \bibfield  {author} {\bibinfo {author} {\bibfnamefont {P.~G.}\ \bibnamefont
  {De~Gennes}},\ }\bibfield  {title} {\bibinfo {title} {{Reptation of a polymer
  chain in the presence of fixed obstacles}},\ }\href
  {https://doi.org/10.1063/1.1675789} {\bibfield  {journal} {\bibinfo
  {journal} {The Journal of Chemical Physics}\ }\textbf {\bibinfo {volume}
  {55}},\ \bibinfo {pages} {572} (\bibinfo {year} {1971})}\BibitemShut
  {NoStop}%
\bibitem [{\citenamefont {Xing}\ \emph {et~al.}(2018)\citenamefont {Xing},
  \citenamefont {Caciagli}, \citenamefont {Cao}, \citenamefont {Stoev},
  \citenamefont {Zupkauskas}, \citenamefont {O’Neill}, \citenamefont
  {Wenzel}, \citenamefont {Lamboll}, \citenamefont {Liu},\ and\ \citenamefont
  {Eiser}}]{Xing2018MicrorheologyHydrogels}%
  \BibitemOpen
  \bibfield  {author} {\bibinfo {author} {\bibfnamefont {Z.}~\bibnamefont
  {Xing}}, \bibinfo {author} {\bibfnamefont {A.}~\bibnamefont {Caciagli}},
  \bibinfo {author} {\bibfnamefont {T.}~\bibnamefont {Cao}}, \bibinfo {author}
  {\bibfnamefont {I.}~\bibnamefont {Stoev}}, \bibinfo {author} {\bibfnamefont
  {M.}~\bibnamefont {Zupkauskas}}, \bibinfo {author} {\bibfnamefont
  {T.}~\bibnamefont {O’Neill}}, \bibinfo {author} {\bibfnamefont
  {T.}~\bibnamefont {Wenzel}}, \bibinfo {author} {\bibfnamefont
  {R.}~\bibnamefont {Lamboll}}, \bibinfo {author} {\bibfnamefont
  {D.}~\bibnamefont {Liu}},\ and\ \bibinfo {author} {\bibfnamefont
  {E.}~\bibnamefont {Eiser}},\ }\bibfield  {title} {\bibinfo {title}
  {{Microrheology of DNA hydrogels}},\ }\href
  {https://doi.org/10.1073/pnas.1722206115} {\bibfield  {journal} {\bibinfo
  {journal} {Proceedings of the National Academy of Sciences}\ }\textbf
  {\bibinfo {volume} {115}},\ \bibinfo {pages} {8137} (\bibinfo {year}
  {2018})}\BibitemShut {NoStop}%
\bibitem [{\citenamefont {Biffi}\ \emph {et~al.}(2013)\citenamefont {Biffi},
  \citenamefont {Cerbino}, \citenamefont {Bomboi}, \citenamefont {Paraboschi},
  \citenamefont {Asselta}, \citenamefont {Sciortino},\ and\ \citenamefont
  {Bellini}}]{Biffi2013PhaseNanostars}%
  \BibitemOpen
  \bibfield  {author} {\bibinfo {author} {\bibfnamefont {S.}~\bibnamefont
  {Biffi}}, \bibinfo {author} {\bibfnamefont {R.}~\bibnamefont {Cerbino}},
  \bibinfo {author} {\bibfnamefont {F.}~\bibnamefont {Bomboi}}, \bibinfo
  {author} {\bibfnamefont {E.~M.}\ \bibnamefont {Paraboschi}}, \bibinfo
  {author} {\bibfnamefont {R.}~\bibnamefont {Asselta}}, \bibinfo {author}
  {\bibfnamefont {F.}~\bibnamefont {Sciortino}},\ and\ \bibinfo {author}
  {\bibfnamefont {T.}~\bibnamefont {Bellini}},\ }\bibfield  {title} {\bibinfo
  {title} {{Phase behavior and critical activated dynamics of limited-valence
  DNA nanostars}},\ }\href {https://doi.org/10.1073/pnas.1304632110} {\bibfield
   {journal} {\bibinfo  {journal} {Proceedings of the National Academy of
  Sciences}\ }\textbf {\bibinfo {volume} {110}},\ \bibinfo {pages} {15633}
  (\bibinfo {year} {2013})}\BibitemShut {NoStop}%
\bibitem [{\citenamefont {Gross}\ \emph {et~al.}(2011)\citenamefont {Gross},
  \citenamefont {Laurens}, \citenamefont {Oddershede}, \citenamefont
  {Bockelmann}, \citenamefont {Peterman},\ and\ \citenamefont
  {Wuite}}]{Gross2011QuantifyingTension}%
  \BibitemOpen
  \bibfield  {author} {\bibinfo {author} {\bibfnamefont {P.}~\bibnamefont
  {Gross}}, \bibinfo {author} {\bibfnamefont {N.}~\bibnamefont {Laurens}},
  \bibinfo {author} {\bibfnamefont {L.~B.}\ \bibnamefont {Oddershede}},
  \bibinfo {author} {\bibfnamefont {U.}~\bibnamefont {Bockelmann}}, \bibinfo
  {author} {\bibfnamefont {E.~J.}\ \bibnamefont {Peterman}},\ and\ \bibinfo
  {author} {\bibfnamefont {G.~J.}\ \bibnamefont {Wuite}},\ }\bibfield  {title}
  {\bibinfo {title} {{Quantifying how DNA stretches, melts and changes twist
  under tension}},\ }\href {https://doi.org/10.1038/nphys2002} {\bibfield
  {journal} {\bibinfo  {journal} {Nature Physics}\ }\textbf {\bibinfo {volume}
  {7}},\ \bibinfo {pages} {731} (\bibinfo {year} {2011})}\BibitemShut {NoStop}%
\bibitem [{\citenamefont {Van~Ketel}\ \emph {et~al.}(2005)\citenamefont
  {Van~Ketel}, \citenamefont {Das},\ and\ \citenamefont
  {Frenkel}}]{VanKetel2005StructuralGas}%
  \BibitemOpen
  \bibfield  {author} {\bibinfo {author} {\bibfnamefont {W.}~\bibnamefont
  {Van~Ketel}}, \bibinfo {author} {\bibfnamefont {C.}~\bibnamefont {Das}},\
  and\ \bibinfo {author} {\bibfnamefont {D.}~\bibnamefont {Frenkel}},\
  }\bibfield  {title} {\bibinfo {title} {{Structural arrest in an ideal gas}},\
  }\href {https://doi.org/10.1103/PhysRevLett.94.135703} {\bibfield  {journal}
  {\bibinfo  {journal} {Physical Review Letters}\ }\textbf {\bibinfo {volume}
  {94}},\ \bibinfo {pages} {8} (\bibinfo {year} {2005})}\BibitemShut {NoStop}%
\bibitem [{\citenamefont {Heine}\ \emph {et~al.}(2010)\citenamefont {Heine},
  \citenamefont {Petersen},\ and\ \citenamefont
  {Grest}}]{Heine2010EffectSuspensions}%
  \BibitemOpen
  \bibfield  {author} {\bibinfo {author} {\bibfnamefont {D.~R.}\ \bibnamefont
  {Heine}}, \bibinfo {author} {\bibfnamefont {M.~K.}\ \bibnamefont
  {Petersen}},\ and\ \bibinfo {author} {\bibfnamefont {G.~S.}\ \bibnamefont
  {Grest}},\ }\bibfield  {title} {\bibinfo {title} {{Effect of particle shape
  and charge on bulk rheology of nanoparticle suspensions}},\ }\bibfield
  {journal} {\bibinfo  {journal} {Journal of Chemical Physics}\ }\textbf
  {\bibinfo {volume} {132}},\ \href {https://doi.org/10.1063/1.3419071}
  {10.1063/1.3419071} (\bibinfo {year} {2010})\BibitemShut {NoStop}%
\bibitem [{\citenamefont {Petersen}\ \emph {et~al.}(2010)\citenamefont
  {Petersen}, \citenamefont {Lane},\ and\ \citenamefont
  {Grest}}]{Petersen2010ShearNanoparticles}%
  \BibitemOpen
  \bibfield  {author} {\bibinfo {author} {\bibfnamefont {M.~K.}\ \bibnamefont
  {Petersen}}, \bibinfo {author} {\bibfnamefont {J.~M.~D.}\ \bibnamefont
  {Lane}},\ and\ \bibinfo {author} {\bibfnamefont {G.~S.}\ \bibnamefont
  {Grest}},\ }\bibfield  {title} {\bibinfo {title} {{Shear rheology of extended
  nanoparticles}},\ }\href {https://doi.org/10.1103/PhysRevE.82.010201}
  {\bibfield  {journal} {\bibinfo  {journal} {Physical Review E - Statistical,
  Nonlinear, and Soft Matter Physics}\ }\textbf {\bibinfo {volume} {82}},\
  \bibinfo {pages} {1} (\bibinfo {year} {2010})}\BibitemShut {NoStop}%
\bibitem [{\citenamefont {King}\ \emph
  {et~al.}(2021{\natexlab{a}})\citenamefont {King}, \citenamefont {Doi},\ and\
  \citenamefont {Eiser}}]{King2021ElasticModelc}%
  \BibitemOpen
  \bibfield  {author} {\bibinfo {author} {\bibfnamefont {D.~A.}\ \bibnamefont
  {King}}, \bibinfo {author} {\bibfnamefont {M.}~\bibnamefont {Doi}},\ and\
  \bibinfo {author} {\bibfnamefont {E.}~\bibnamefont {Eiser}},\ }\bibfield
  {title} {\bibinfo {title} {{Elastic response of wire frame glasses. I. Two
  dimensional model}},\ }\href {https://doi.org/10.1063/5.0046524} {\bibfield
  {journal} {\bibinfo  {journal} {The Journal of Chemical Physics}\ }\textbf
  {\bibinfo {volume} {154}},\ \bibinfo {pages} {244904} (\bibinfo {year}
  {2021}{\natexlab{a}})}\BibitemShut {NoStop}%
\bibitem [{\citenamefont {King}\ \emph
  {et~al.}(2021{\natexlab{b}})\citenamefont {King}, \citenamefont {Doi},\ and\
  \citenamefont {Eiser}}]{King2021ElasticSystemsc}%
  \BibitemOpen
  \bibfield  {author} {\bibinfo {author} {\bibfnamefont {D.~A.}\ \bibnamefont
  {King}}, \bibinfo {author} {\bibfnamefont {M.}~\bibnamefont {Doi}},\ and\
  \bibinfo {author} {\bibfnamefont {E.}~\bibnamefont {Eiser}},\ }\bibfield
  {title} {\bibinfo {title} {{Elastic response of wire frame glasses. II.
  Three-dimensional systems}},\ }\href {https://doi.org/10.1063/5.0046525}
  {\bibfield  {journal} {\bibinfo  {journal} {The Journal of Chemical Physics}\
  }\textbf {\bibinfo {volume} {154}},\ \bibinfo {pages} {244905} (\bibinfo
  {year} {2021}{\natexlab{b}})}\BibitemShut {NoStop}%
\bibitem [{\citenamefont {Yardimci}\ \emph {et~al.}(2023)\citenamefont
  {Yardimci}, \citenamefont {Gibaud}, \citenamefont {Schwenger}, \citenamefont
  {Sartucci}, \citenamefont {Olmsted}, \citenamefont {Urbach},\ and\
  \citenamefont {Dogic}}]{Yardimci2023}%
  \BibitemOpen
  \bibfield  {author} {\bibinfo {author} {\bibfnamefont {S.}~\bibnamefont
  {Yardimci}}, \bibinfo {author} {\bibfnamefont {T.}~\bibnamefont {Gibaud}},
  \bibinfo {author} {\bibfnamefont {W.}~\bibnamefont {Schwenger}}, \bibinfo
  {author} {\bibfnamefont {M.~R.}\ \bibnamefont {Sartucci}}, \bibinfo {author}
  {\bibfnamefont {P.~D.}\ \bibnamefont {Olmsted}}, \bibinfo {author}
  {\bibfnamefont {J.~S.}\ \bibnamefont {Urbach}},\ and\ \bibinfo {author}
  {\bibfnamefont {Z.}~\bibnamefont {Dogic}},\ }\bibfield  {title} {\bibinfo
  {title} {Bonded straight and helical flagellar filaments form
  ultra-low-density glasses},\ }\bibfield  {journal} {\bibinfo  {journal}
  {Proceedings of the National Academy of Sciences}\ }\textbf {\bibinfo
  {volume} {120}},\ \href {https://doi.org/10.1073/pnas.2215766120}
  {10.1073/pnas.2215766120} (\bibinfo {year} {2023})\BibitemShut {NoStop}%
\bibitem [{\citenamefont {Cichocki}(1987)}]{Cichocki1987TheCores}%
  \BibitemOpen
  \bibfield  {author} {\bibinfo {author} {\bibfnamefont {B.}~\bibnamefont
  {Cichocki}},\ }\bibfield  {title} {\bibinfo {title} {{The generalized
  Smoluchowski equation for interacting Brownian particles with hard cores}},\
  }\href {https://doi.org/10.1007/BF01303903} {\bibfield  {journal} {\bibinfo
  {journal} {Zeitschrift f{\"{u}}r Physik B Condensed Matter}\ }\textbf
  {\bibinfo {volume} {66}},\ \bibinfo {pages} {537} (\bibinfo {year}
  {1987})}\BibitemShut {NoStop}%
\bibitem [{\citenamefont {Schilling}\ and\ \citenamefont
  {Szamel}(2003)}]{Schilling2003MicroscopicCorrelations}%
  \BibitemOpen
  \bibfield  {author} {\bibinfo {author} {\bibfnamefont {R.}~\bibnamefont
  {Schilling}}\ and\ \bibinfo {author} {\bibfnamefont {G.}~\bibnamefont
  {Szamel}},\ }\bibfield  {title} {\bibinfo {title} {{Microscopic theory for
  the glass transition in a system without static correlations}},\ }\href
  {https://doi.org/10.1088/0953-8984/15/11/320} {\bibfield  {journal} {\bibinfo
   {journal} {Europhysics Letters}\ }\textbf {\bibinfo {volume} {61}},\
  \bibinfo {pages} {207} (\bibinfo {year} {2003})}\BibitemShut {NoStop}%
\bibitem [{\citenamefont {Sussman}\ and\ \citenamefont
  {Schweizer}(2011)}]{Sussman2011MicroscopicMacromolecules}%
  \BibitemOpen
  \bibfield  {author} {\bibinfo {author} {\bibfnamefont {D.~M.}\ \bibnamefont
  {Sussman}}\ and\ \bibinfo {author} {\bibfnamefont {K.~S.}\ \bibnamefont
  {Schweizer}},\ }\bibfield  {title} {\bibinfo {title} {{Microscopic theory of
  topologically entangled fluids of rigid macromolecules}},\ }\href
  {https://doi.org/10.1103/PhysRevE.83.061501} {\bibfield  {journal} {\bibinfo
  {journal} {Physical Review E - Statistical, Nonlinear, and Soft Matter
  Physics}\ }\textbf {\bibinfo {volume} {83}},\ \bibinfo {pages} {1} (\bibinfo
  {year} {2011})}\BibitemShut {NoStop}%
\bibitem [{\citenamefont {Szamel}(1993)}]{Szamel1993ReptationPolymers}%
  \BibitemOpen
  \bibfield  {author} {\bibinfo {author} {\bibfnamefont {G.}~\bibnamefont
  {Szamel}},\ }\bibfield  {title} {\bibinfo {title} {{Reptation as a dynamic
  mean-field theory: Study of a simple model of rodlike polymers}},\ }\href
  {https://doi.org/10.1103/PhysRevLett.70.3744} {\bibfield  {journal} {\bibinfo
   {journal} {Physical Review Letters}\ }\textbf {\bibinfo {volume} {70}},\
  \bibinfo {pages} {3744} (\bibinfo {year} {1993})}\BibitemShut {NoStop}%
\bibitem [{\citenamefont {Szamel}\ and\ \citenamefont
  {Schweizer}(1994)}]{Szamel1994ReptationPolymers}%
  \BibitemOpen
  \bibfield  {author} {\bibinfo {author} {\bibfnamefont {G.}~\bibnamefont
  {Szamel}}\ and\ \bibinfo {author} {\bibfnamefont {K.~S.}\ \bibnamefont
  {Schweizer}},\ }\bibfield  {title} {\bibinfo {title} {{Reptation as a dynamic
  mean-field theory: Self and tracer diffusion in a simple model of rodlike
  polymers}},\ }\href {https://doi.org/10.1063/1.466403} {\bibfield  {journal}
  {\bibinfo  {journal} {The Journal of Chemical Physics}\ }\textbf {\bibinfo
  {volume} {100}},\ \bibinfo {pages} {3127} (\bibinfo {year}
  {1994})}\BibitemShut {NoStop}%
\bibitem [{\citenamefont {Reichman}\ and\ \citenamefont
  {Charbonneau}(2005)}]{Reichman2005}%
  \BibitemOpen
  \bibfield  {author} {\bibinfo {author} {\bibfnamefont {D.~R.}\ \bibnamefont
  {Reichman}}\ and\ \bibinfo {author} {\bibfnamefont {P.}~\bibnamefont
  {Charbonneau}},\ }\bibfield  {title} {\bibinfo {title} {{Mode-coupling
  theory}},\ }\href {https://doi.org/10.1088/1742-5468/2005/05/P05013}
  {\bibfield  {journal} {\bibinfo  {journal} {Journal of Statistical Mechanics:
  Theory and Experiment}\ ,\ \bibinfo {pages} {267}} (\bibinfo {year}
  {2005})}\BibitemShut {NoStop}%
\bibitem [{\citenamefont {Fuchs}\ and\ \citenamefont
  {Cates}(2009)}]{Fuchs2009AFlow}%
  \BibitemOpen
  \bibfield  {author} {\bibinfo {author} {\bibfnamefont {M.}~\bibnamefont
  {Fuchs}}\ and\ \bibinfo {author} {\bibfnamefont {M.~E.}\ \bibnamefont
  {Cates}},\ }\bibfield  {title} {\bibinfo {title} {{A mode coupling theory for
  Brownian particles in homogeneous steady shear flow}},\ }\href
  {https://doi.org/10.1122/1.3119084} {\bibfield  {journal} {\bibinfo
  {journal} {Journal of Rheology}\ }\textbf {\bibinfo {volume} {53}},\ \bibinfo
  {pages} {957} (\bibinfo {year} {2009})}\BibitemShut {NoStop}%
\bibitem [{\citenamefont {Miyazaki}\ and\ \citenamefont
  {Yethiraj}(2002)}]{Miyazaki2002}%
  \BibitemOpen
  \bibfield  {author} {\bibinfo {author} {\bibfnamefont {K.}~\bibnamefont
  {Miyazaki}}\ and\ \bibinfo {author} {\bibfnamefont {A.}~\bibnamefont
  {Yethiraj}},\ }\bibfield  {title} {\bibinfo {title} {{Entanglement effects in
  mode coupling theories of polymers}},\ }\href
  {https://doi.org/10.1063/1.1527943} {\bibfield  {journal} {\bibinfo
  {journal} {Journal of Chemical Physics}\ }\textbf {\bibinfo {volume} {117}},\
  \bibinfo {pages} {10448} (\bibinfo {year} {2002})}\BibitemShut {NoStop}%
\bibitem [{\citenamefont {Edwards}\ and\ \citenamefont
  {Evans}(1982)}]{Edwards1982DynamicsMolecules}%
  \BibitemOpen
  \bibfield  {author} {\bibinfo {author} {\bibfnamefont {S.~F.}\ \bibnamefont
  {Edwards}}\ and\ \bibinfo {author} {\bibfnamefont {K.~E.}\ \bibnamefont
  {Evans}},\ }\bibfield  {title} {\bibinfo {title} {{Dynamics of highly
  entangled rod-like molecules}},\ }\href
  {https://doi.org/10.1039/F29827800113} {\bibfield  {journal} {\bibinfo
  {journal} {Journal of the Chemical Society, Faraday Transactions 2: Molecular
  and Chemical Physics}\ }\textbf {\bibinfo {volume} {78}},\ \bibinfo {pages}
  {113} (\bibinfo {year} {1982})}\BibitemShut {NoStop}%
\bibitem [{\citenamefont {Edwards}\ and\ \citenamefont
  {Vilgis}(1986)}]{Edwards1986TheTransition}%
  \BibitemOpen
  \bibfield  {author} {\bibinfo {author} {\bibfnamefont {S.~F.}\ \bibnamefont
  {Edwards}}\ and\ \bibinfo {author} {\bibfnamefont {T.}~\bibnamefont
  {Vilgis}},\ }\bibfield  {title} {\bibinfo {title} {{The dynamics of the glass
  transition}},\ }\href {https://doi.org/10.1088/0031-8949/1986/T13/001}
  {\bibfield  {journal} {\bibinfo  {journal} {Physica Scripta}\ }\textbf
  {\bibinfo {volume} {1986}},\ \bibinfo {pages} {7} (\bibinfo {year}
  {1986})}\BibitemShut {NoStop}%
\bibitem [{\citenamefont {Teraoka}\ and\ \citenamefont
  {Hayakawa}(1988)}]{Teraoka1988TheoryDiffusion}%
  \BibitemOpen
  \bibfield  {author} {\bibinfo {author} {\bibfnamefont {I.}~\bibnamefont
  {Teraoka}}\ and\ \bibinfo {author} {\bibfnamefont {R.}~\bibnamefont
  {Hayakawa}},\ }\bibfield  {title} {\bibinfo {title} {{Theory of dynamics of
  entangled rod-like polymers by use of a mean-field Green function
  formulation. I. Transverse diffusion}},\ }\href
  {https://doi.org/10.1063/1.455325} {\bibfield  {journal} {\bibinfo  {journal}
  {The Journal of Chemical Physics}\ }\textbf {\bibinfo {volume} {89}},\
  \bibinfo {pages} {6989} (\bibinfo {year} {1988})}\BibitemShut {NoStop}%
\bibitem [{\citenamefont {Teraoka}\ and\ \citenamefont
  {Hayakawa}(1989)}]{Teraoka1989}%
  \BibitemOpen
  \bibfield  {author} {\bibinfo {author} {\bibfnamefont {I.}~\bibnamefont
  {Teraoka}}\ and\ \bibinfo {author} {\bibfnamefont {R.}~\bibnamefont
  {Hayakawa}},\ }\bibfield  {title} {\bibinfo {title} {{Theory of dynamics of
  entangled rod‐like polymers by use of a mean‐field Green function
  formulation. II. Rotational diffusion}},\ }\href
  {https://doi.org/10.1063/1.456973} {\bibfield  {journal} {\bibinfo  {journal}
  {The Journal of Chemical Physics}\ }\textbf {\bibinfo {volume} {91}},\
  \bibinfo {pages} {2643} (\bibinfo {year} {1989})}\BibitemShut {NoStop}%
\bibitem [{\citenamefont {Teraoka}\ and\ \citenamefont
  {Karasz}(1992)}]{Teraoka1992One-dimensionalBarriers}%
  \BibitemOpen
  \bibfield  {author} {\bibinfo {author} {\bibfnamefont {I.}~\bibnamefont
  {Teraoka}}\ and\ \bibinfo {author} {\bibfnamefont {F.~E.}\ \bibnamefont
  {Karasz}},\ }\bibfield  {title} {\bibinfo {title} {{One-dimensional diffusion
  with stochastic random barriers}},\ }\href
  {https://doi.org/10.1103/PhysRevA.45.5426} {\bibfield  {journal} {\bibinfo
  {journal} {Physical Review A}\ }\textbf {\bibinfo {volume} {45}},\ \bibinfo
  {pages} {5426} (\bibinfo {year} {1992})}\BibitemShut {NoStop}%
\bibitem [{\citenamefont {Teraoka}\ and\ \citenamefont
  {Karasz}(1993)}]{Teraoka1993GlassMolecules}%
  \BibitemOpen
  \bibfield  {author} {\bibinfo {author} {\bibfnamefont {I.}~\bibnamefont
  {Teraoka}}\ and\ \bibinfo {author} {\bibfnamefont {F.~E.}\ \bibnamefont
  {Karasz}},\ }\bibfield  {title} {\bibinfo {title} {{Glass transition and
  dynamic-mobility spectrum of an isotropic system of rodlike molecules}},\
  }\href {https://doi.org/10.1103/PhysRevE.47.1108} {\bibfield  {journal}
  {\bibinfo  {journal} {Physical Review E}\ }\textbf {\bibinfo {volume} {47}},\
  \bibinfo {pages} {1108} (\bibinfo {year} {1993})}\BibitemShut {NoStop}%
\bibitem [{\citenamefont {Bouchaud}\ \emph {et~al.}(1996)\citenamefont
  {Bouchaud}, \citenamefont {Cugliandolo}, \citenamefont {Kurchan},\ and\
  \citenamefont {M{\'{e}}zard}}]{Bouchaud1996}%
  \BibitemOpen
  \bibfield  {author} {\bibinfo {author} {\bibfnamefont {J.-P.}\ \bibnamefont
  {Bouchaud}}, \bibinfo {author} {\bibfnamefont {L.}~\bibnamefont
  {Cugliandolo}}, \bibinfo {author} {\bibfnamefont {J.}~\bibnamefont
  {Kurchan}},\ and\ \bibinfo {author} {\bibfnamefont {M.}~\bibnamefont
  {M{\'{e}}zard}},\ }\bibfield  {title} {\bibinfo {title} {{Mode-coupling
  approximations, glass theory and disordered systems}},\ }\href
  {https://doi.org/10.1016/0378-4371(95)00423-8} {\bibfield  {journal}
  {\bibinfo  {journal} {Physica A: Statistical Mechanics and its Applications}\
  }\textbf {\bibinfo {volume} {226}},\ \bibinfo {pages} {243} (\bibinfo {year}
  {1996})}\BibitemShut {NoStop}%
\bibitem [{\citenamefont {Frenkel}\ and\ \citenamefont
  {Maguire}(1983)}]{Frenkel1983MolecularRods}%
  \BibitemOpen
  \bibfield  {author} {\bibinfo {author} {\bibfnamefont {D.}~\bibnamefont
  {Frenkel}}\ and\ \bibinfo {author} {\bibfnamefont {J.~F.}\ \bibnamefont
  {Maguire}},\ }\bibfield  {title} {\bibinfo {title} {{Molecular dynamics study
  of the dynamical properties an assembly of infinitely thin hard rods}},\
  }\href@noop {} {\bibfield  {journal} {\bibinfo  {journal} {Molecular
  Physics}\ }\textbf {\bibinfo {volume} {49}},\ \bibinfo {pages} {503}
  (\bibinfo {year} {1983})}\BibitemShut {NoStop}%
\bibitem [{\citenamefont {Frenkel}\ and\ \citenamefont
  {Maguire}(1981)}]{Frenkel1981MolecularProperties}%
  \BibitemOpen
  \bibfield  {author} {\bibinfo {author} {\bibfnamefont {D.}~\bibnamefont
  {Frenkel}}\ and\ \bibinfo {author} {\bibfnamefont {J.~F.}\ \bibnamefont
  {Maguire}},\ }\bibfield  {title} {\bibinfo {title} {{Molecular Dynamics Study
  of Infinitely Thin Hard Rods: Scaling Behavior of Transport Properties}},\
  }\href {https://doi.org/10.1103/PhysRevLett.47.1025} {\bibfield  {journal}
  {\bibinfo  {journal} {Physical Review Letters}\ }\textbf {\bibinfo {volume}
  {47}},\ \bibinfo {pages} {1025} (\bibinfo {year} {1981})}\BibitemShut
  {NoStop}%
\bibitem [{\citenamefont {Edwards}(1965)}]{Edwards1965ExactCoefficients}%
  \BibitemOpen
  \bibfield  {author} {\bibinfo {author} {\bibfnamefont {S.~F.}\ \bibnamefont
  {Edwards}},\ }\bibfield  {title} {\bibinfo {title} {{Exact formulae for
  inverse transport coefficients}},\ }\href
  {https://doi.org/10.1088/0370-1328/86/5/309} {\bibfield  {journal} {\bibinfo
  {journal} {Proceedings of the Physical Society}\ }\textbf {\bibinfo {volume}
  {86}},\ \bibinfo {pages} {977} (\bibinfo {year} {1965})}\BibitemShut
  {NoStop}%
\bibitem [{\citenamefont {King}\ \emph {et~al.}(2020)\citenamefont {King},
  \citenamefont {Doi},\ and\ \citenamefont
  {Eiser}}]{King2020ParticleSuspensionsb}%
  \BibitemOpen
  \bibfield  {author} {\bibinfo {author} {\bibfnamefont {D.~A.}\ \bibnamefont
  {King}}, \bibinfo {author} {\bibfnamefont {M.}~\bibnamefont {Doi}},\ and\
  \bibinfo {author} {\bibfnamefont {E.}~\bibnamefont {Eiser}},\ }\bibfield
  {title} {\bibinfo {title} {{Particle shapes leading to Newtonian dilute
  suspensions}},\ }\href {https://doi.org/10.1103/PhysRevE.102.032615}
  {\bibfield  {journal} {\bibinfo  {journal} {Physical Review E}\ }\textbf
  {\bibinfo {volume} {102}},\ \bibinfo {pages} {032615} (\bibinfo {year}
  {2020})}\BibitemShut {NoStop}%
\bibitem [{\citenamefont {King}(2021)}]{King2021EffectsDynamics}%
  \BibitemOpen
  \bibfield  {author} {\bibinfo {author} {\bibfnamefont {D.~A.}\ \bibnamefont
  {King}},\ }\emph {\bibinfo {title} {{Effects of particle shape and
  flexibility on suspension dynamics}}},\ \href
  {https://doi.org/10.17863/CAM.79382} {Ph.D. thesis},\ \bibinfo  {school}
  {University of Cambridge} (\bibinfo {year} {2021})\BibitemShut {NoStop}%
\bibitem [{\citenamefont {Kirkwood}(1949)}]{Kirkwood1949TheBehavior}%
  \BibitemOpen
  \bibfield  {author} {\bibinfo {author} {\bibfnamefont {J.~G.}\ \bibnamefont
  {Kirkwood}},\ }\bibfield  {title} {\bibinfo {title} {{The statistical
  mechanical theory of irreversible processes in solutions of flexible
  macromolecules. Visco-elastic behavior}},\ }\href
  {https://doi.org/10.1002/recl.19490680708} {\bibfield  {journal} {\bibinfo
  {journal} {Recueil des Travaux Chimiques des Pays-Bas}\ }\textbf {\bibinfo
  {volume} {68}},\ \bibinfo {pages} {649} (\bibinfo {year} {1949})}\BibitemShut
  {NoStop}%
\bibitem [{\citenamefont {Kirkwood}(1954)}]{Kirkwood1954TheMacromolecules}%
  \BibitemOpen
  \bibfield  {author} {\bibinfo {author} {\bibfnamefont {J.~G.}\ \bibnamefont
  {Kirkwood}},\ }\bibfield  {title} {\bibinfo {title} {{The general theory of
  irreversible processes in solutions of macromolecules}},\ }\href
  {https://doi.org/10.1002/polb.1996.897} {\bibfield  {journal} {\bibinfo
  {journal} {Journal of Polymer Science Part B: Polymer Physics}\ }\textbf
  {\bibinfo {volume} {12}},\ \bibinfo {pages} {1} (\bibinfo {year}
  {1954})}\BibitemShut {NoStop}%
\bibitem [{\citenamefont {Fixman}(1991)}]{Fixman1991StressSolutions}%
  \BibitemOpen
  \bibfield  {author} {\bibinfo {author} {\bibfnamefont {M.}~\bibnamefont
  {Fixman}},\ }\bibfield  {title} {\bibinfo {title} {{Stress relaxation in
  polymer melts and concentrated solutions}},\ }\href
  {https://doi.org/10.1063/1.461808} {\bibfield  {journal} {\bibinfo  {journal}
  {The Journal of Chemical Physics}\ }\textbf {\bibinfo {volume} {95}},\
  \bibinfo {pages} {1410} (\bibinfo {year} {1991})}\BibitemShut {NoStop}%
\bibitem [{\citenamefont {Muthukumar}\ and\ \citenamefont
  {Edwards}(1983)}]{Muthukumar1983ScreeningMacromolecules}%
  \BibitemOpen
  \bibfield  {author} {\bibinfo {author} {\bibfnamefont {M.}~\bibnamefont
  {Muthukumar}}\ and\ \bibinfo {author} {\bibfnamefont {S.~F.}\ \bibnamefont
  {Edwards}},\ }\bibfield  {title} {\bibinfo {title} {{Screening of
  Hydrodynamic Interaction in a Solution of Rodlike Macromolecules}},\
  }\href@noop {} {\bibfield  {journal} {\bibinfo  {journal} {Macromolecules}\
  }\textbf {\bibinfo {volume} {16}},\ \bibinfo {pages} {1475} (\bibinfo {year}
  {1983})}\BibitemShut {NoStop}%
\bibitem [{\citenamefont {Edwards}\ and\ \citenamefont
  {Freed}(1974)}]{Edwards1974TheorySolutions}%
  \BibitemOpen
  \bibfield  {author} {\bibinfo {author} {\bibfnamefont {S.~F.}\ \bibnamefont
  {Edwards}}\ and\ \bibinfo {author} {\bibfnamefont {K.~F.}\ \bibnamefont
  {Freed}},\ }\bibfield  {title} {\bibinfo {title} {{Theory of the dynamical
  viscosity of polymer solutions}},\ }\href {https://doi.org/10.1063/1.1681993}
  {\bibfield  {journal} {\bibinfo  {journal} {The Journal of Chemical Physics}\
  }\textbf {\bibinfo {volume} {61}},\ \bibinfo {pages} {1189} (\bibinfo {year}
  {1974})}\BibitemShut {NoStop}%
\bibitem [{\citenamefont {Freed}\ and\ \citenamefont
  {Edwards}(1974)}]{Freed1974PolymerSolutions}%
  \BibitemOpen
  \bibfield  {author} {\bibinfo {author} {\bibfnamefont {K.~F.}\ \bibnamefont
  {Freed}}\ and\ \bibinfo {author} {\bibfnamefont {S.~F.}\ \bibnamefont
  {Edwards}},\ }\bibfield  {title} {\bibinfo {title} {{Polymer viscosity in
  concentrated solutions}},\ }\href {https://doi.org/10.1063/1.1682545}
  {\bibfield  {journal} {\bibinfo  {journal} {The Journal of Chemical Physics}\
  }\textbf {\bibinfo {volume} {61}},\ \bibinfo {pages} {3626} (\bibinfo {year}
  {1974})}\BibitemShut {NoStop}%
\bibitem [{\citenamefont {Fredrickson}\ and\ \citenamefont
  {Shaqfeh}(1989)}]{Fredrickson1989HeatFibers}%
  \BibitemOpen
  \bibfield  {author} {\bibinfo {author} {\bibfnamefont {G.~H.}\ \bibnamefont
  {Fredrickson}}\ and\ \bibinfo {author} {\bibfnamefont {E.~S.}\ \bibnamefont
  {Shaqfeh}},\ }\bibfield  {title} {\bibinfo {title} {{Heat and mass transport
  in composites of aligned slender fibers}},\ }\href
  {https://doi.org/10.1063/1.857546} {\bibfield  {journal} {\bibinfo  {journal}
  {Physics of Fluids A}\ }\textbf {\bibinfo {volume} {1}},\ \bibinfo {pages}
  {3} (\bibinfo {year} {1989})}\BibitemShut {NoStop}%
\bibitem [{\citenamefont {Shaqfeh}\ and\ \citenamefont
  {Fredrickson}(1990)}]{Shaqfeh1990TheRods}%
  \BibitemOpen
  \bibfield  {author} {\bibinfo {author} {\bibfnamefont {E.~S.~G.}\
  \bibnamefont {Shaqfeh}}\ and\ \bibinfo {author} {\bibfnamefont {G.~H.}\
  \bibnamefont {Fredrickson}},\ }\bibfield  {title} {\bibinfo {title} {{The
  hydrodynamic stress in a suspension of rods}},\ }\href
  {https://doi.org/10.1063/1.857683} {\bibfield  {journal} {\bibinfo  {journal}
  {Physics of Fluids A: Fluid Dynamics}\ }\textbf {\bibinfo {volume} {2}},\
  \bibinfo {pages} {7} (\bibinfo {year} {1990})}\BibitemShut {NoStop}%
\bibitem [{\citenamefont {Bixon}\ and\ \citenamefont
  {Zwanzig}(1981)}]{Bixon1981DiffusionTraps}%
  \BibitemOpen
  \bibfield  {author} {\bibinfo {author} {\bibfnamefont {M.}~\bibnamefont
  {Bixon}}\ and\ \bibinfo {author} {\bibfnamefont {R.}~\bibnamefont
  {Zwanzig}},\ }\bibfield  {title} {\bibinfo {title} {{Diffusion in a medium
  with static traps}},\ }\href {https://doi.org/10.1063/1.442297} {\bibfield
  {journal} {\bibinfo  {journal} {The Journal of Chemical Physics}\ }\textbf
  {\bibinfo {volume} {75}},\ \bibinfo {pages} {2354} (\bibinfo {year}
  {1981})}\BibitemShut {NoStop}%
\bibitem [{\citenamefont {Kirkpatrick}(1982)}]{Kirkpatrick1982TimeTraps}%
  \BibitemOpen
  \bibfield  {author} {\bibinfo {author} {\bibfnamefont {T.~R.}\ \bibnamefont
  {Kirkpatrick}},\ }\bibfield  {title} {\bibinfo {title} {{Time dependent
  transport in a fluid with static traps}},\ }\href
  {https://doi.org/10.1063/1.443503} {\bibfield  {journal} {\bibinfo  {journal}
  {The Journal of Chemical Physics}\ }\textbf {\bibinfo {volume} {76}},\
  \bibinfo {pages} {4255} (\bibinfo {year} {1982})}\BibitemShut {NoStop}%
\bibitem [{\citenamefont {Grassberger}\ and\ \citenamefont
  {Procaccia}(1982)}]{Grassberger1982TheTraps}%
  \BibitemOpen
  \bibfield  {author} {\bibinfo {author} {\bibfnamefont {P.}~\bibnamefont
  {Grassberger}}\ and\ \bibinfo {author} {\bibfnamefont {I.}~\bibnamefont
  {Procaccia}},\ }\bibfield  {title} {\bibinfo {title} {{The long time
  properties of diffusion in a medium with static traps}},\ }\href
  {https://doi.org/10.1063/1.443832} {\bibfield  {journal} {\bibinfo  {journal}
  {The Journal of Chemical Physics}\ }\textbf {\bibinfo {volume} {77}},\
  \bibinfo {pages} {6281} (\bibinfo {year} {1982})}\BibitemShut {NoStop}%
\bibitem [{\citenamefont {Abramowitz}\ and\ \citenamefont
  {Stegun}(1964)}]{Abramowitz1964HandbookTables}%
  \BibitemOpen
  \bibfield  {author} {\bibinfo {author} {\bibfnamefont {M.}~\bibnamefont
  {Abramowitz}}\ and\ \bibinfo {author} {\bibfnamefont {I.~A.}\ \bibnamefont
  {Stegun}},\ }\href@noop {} {\emph {\bibinfo {title} {{Handbook of
  Mathematical Functions with Formulas, Graphs, and Mathematical Tables}}}}\
  (\bibinfo  {publisher} {Dover Publications},\ \bibinfo {address} {New York
  City},\ \bibinfo {year} {1964})\BibitemShut {NoStop}%
\bibitem [{\citenamefont {Copson}(1965)}]{Copson1965}%
  \BibitemOpen
  \bibfield  {author} {\bibinfo {author} {\bibfnamefont {E.~T.}\ \bibnamefont
  {Copson}},\ }\href {https://doi.org/10.1017/CBO9780511526121} {\emph
  {\bibinfo {title} {Asymptotic Expansions}}},\ edited by\ \bibinfo {editor}
  {\bibfnamefont {F.}~\bibnamefont {Smithies}}\ and\ \bibinfo {editor}
  {\bibfnamefont {J.~A.}\ \bibnamefont {Todd}}\ (\bibinfo  {publisher}
  {Cambridge University Press},\ \bibinfo {year} {1965})\BibitemShut {NoStop}%
\end{thebibliography}%

\end{document}